%% file: journal_arxiv.tex
\documentclass[10pt,journal]{IEEEtran}

\usepackage{cite}
\usepackage[pdftex]{graphicx}
\usepackage{amsmath}
\usepackage{amsfonts}
\usepackage{algorithmic}
\usepackage{multirow}
\usepackage[ruled,vlined]{algorithm2e}
\usepackage{array}
\usepackage{caption}
\usepackage{subcaption}
\usepackage{color}
\usepackage{tikz}

\usepackage{booktabs}
\usepackage{footnote}
\makesavenoteenv{table}
\makesavenoteenv{tabular}
\usepackage{siunitx}

\usepackage{fixltx2e}
\usepackage{dblfloatfix}
\usepackage{url}

\usepackage[acronyms,nonumberlist,nopostdot,nomain,nogroupskip]{glossaries}
\input{acronyms.tex}

\makeatother

\newcommand{\edit}[1]{\textcolor{black}{#1}}

\IEEEoverridecommandlockouts
\newcommand\copyrightnotice{%
	\begin{tikzpicture}[remember picture,overlay]
	\node[anchor=north,yshift=-10pt] at (current page.north) {\fbox{\parbox{\dimexpr\textwidth-\fboxsep-\fboxrule\relax}{\footnotesize \textcopyright 2021 IEEE. This paper is under review at IEEE Transaction on Networking. Personal use of this material is permitted. Permission from IEEE must be obtained for all other uses, in any current or future media, including reprinting/republishing this material for advertising or promotional purposes, creating new collective works, for resale or redistribution to servers or lists,	or reuse of any copyrighted component of this work in other works.}}};
	\end{tikzpicture}
}

\begin{document}

\title{Using Distributed Reinforcement Learning for Resource Orchestration in a Network Slicing Scenario
\thanks{This work was supported by Consortium GARR through the ``Orio Carlini" scholarship 2019. A preliminary and reduced version of this manuscript has been submitted to the \textit{IEEE Mediterranean Communication and Computer Networking Conference}, June 2021.}
}

\author{
    
    \IEEEauthorblockN{\large Federico Mason$^*$, Gianfranco Nencioni$^{\dagger}$, Andrea Zanella$^*$ \vspace{1mm}}

    \IEEEauthorblockA{
    	\small $^*$\small\texttt{\{masonfed, zanella\}@dei.unipd.it},
    	\small $^\dagger$\small\texttt{gianfranco.nencioni@uis.no}
    } \\

	\IEEEauthorblockA{
		\small $^*$ Department of Information Engineering, University of Padova - Padova, Italy
	}

	\IEEEauthorblockA{
		\small $^\dagger$ Department of Electrical Engineering and Computer Science, University of Stavanger - Stavanger, Norway
	}
}

\maketitle

\begin{abstract}
	The Network Slicing (NS) paradigm enables the partition of physical and virtual resources among multiple logical networks, possibly managed by different tenants.
	In such a scenario, network resources need to be dynamically allocated according to the slices' requirements.
	In this paper, we attack the above problem by exploiting a Deep Reinforcement Learning approach.
	Our framework is based on a distributed architecture, where multiple agents cooperate towards a common goal.
	The agents' training is carried out following the Advantage Actor Critic algorithm, which allows to handle continuous action spaces. 
	By means of extensive simulations, we show that our approach yields better performance than both a static allocation of system resources and an efficient empirical strategy.
	At the same time, the proposed system ensures high adaptability to different scenarios without the need for additional training.
\end{abstract}


\IEEEpeerreviewmaketitle

\copyrightnotice

\section{Introduction}\label{sec:introduction}

\IEEEPARstart{T}{he} fifth generation of cellular networks (5G) aims at supporting different applications with very specific requirements over the same infrastructure.
In this perspective, the 3GPP consortium has identified three main service classes, namely \gls{embb}, \gls{urllc} and \gls{mmtc}~\cite{3gpp.22.261}.
Specifically, \gls{embb} is expected to provide very high throughput in both downlink and uplink, while \gls{urllc} appeals to applications with strict latency and reliability constraints, like robotic surgeries, tactile Internet, and emergency communications~\cite{navarro2020}.

Traditional telecommunication networks are often based on a rigid architecture and are not apt to support such services~\cite{yang2015}.
To overcome this problem, the research community has introduced the concepts of \gls{sdn} and \gls{nfv}, which can make networks more flexible and adaptable to different requirements~\cite{zhang2017}.
The \gls{nfv} principle makes it possible to execute network functions over virtual machines in the Cloud.
Instead, \gls{sdn} separates the control plane from the forwarding plane, enabling the dynamic routing of data flows. 

In particular, the \gls{sdn} and \gls{nfv} concepts are key enablers of the \gls{ns} paradigm, which makes it possible to define multiple virtual networks over the same physical infrastructure~\cite{rost2017, afolabi2018}.
Under this vision, a slice consists of a virtual overlay network designed to support communication services with similar characteristics~\cite{popovski2018}.
Hence, a slice supporting \gls{embb} applications (e.g., video streaming) is characterized by very high bit rate, while a slice supporting \gls{urllc} applications (e.g., telesurgery) guarantees extremely high reliability and low latency.

If defined over the same infrastructure, different slices will contend for the same resources, which can be both physical (e.g., optical links) and virtual (e.g., virtual baseband processing units)~\cite{richart2016}.
In general, such resources are acquired by the \textit{slice broker} (i.e., the body in charge of initializing and orchestrating slices) from the \textit{infrastructure providers} (i.e., the owners of the physical elements of the network)~\cite{caballero2017}. 
Then, slices are assigned to the \textit{slice tenants} (e.g., virtual network operators), which offer slice services to the end-users. 
The amount of resources that are needed to support the slice services are determined by the so-called \gls{sla} between the slice tenant and broker~\cite{samdanis2016}. 
Therefore, a fundamental challenge in \gls{ns} systems is how to distribute resources among the different slices in an efficient way, ensuring that all the \glspl{sla} are satisfied~\cite{trivisonno2015}.

In this work, we consider a scenario where two slice classes (i.e., \gls{embb} and \gls{urllc}), with dynamic requirements in term of throughput, computational power, memory capacity, and delay, are instantiated over the same network infrastructure. 
Each slice is composed by multiple information flows (with static routes) that contend for the bandwidth provided by network links, and the computational and memory resources provided by the computing facilities connected with the network nodes.
The problem is to dynamically distribute the network resources among the active information flows, in accordance to the characteristic of the slices which they belong to. 

The \emph{naive} approach consists in statically allocating communication and computational resources to the different slices.
However, this method cannot exploit the statistical multiplexing of the information flows and, consequently, may lead to greater over-provisioning costs and low utilization of the available resources.
On the other hand, conventional resource allocation strategies are often unsuitable because cannot understand the specific features of different slices, neither deal with the high complexity of \gls{ns} environments.

Here, we propose a machine-learning based approach and attack the problem by exploiting the \gls{drl} paradigm, which combines \gls{rl} algorithms and \glspl{nn} to find strategies for the management of complex environments~\cite{mnih2015human}.
More specifically, we design a distributed \gls{drl} system, where multiple agents collaborate to allocate network resources among the different slices running over the same infrastructure.
The continuous interaction between such learning units makes it possible to increase both the system efficiency and adaptability to different scenarios.
The main contributions of our work consists in the following points:
\begin{itemize}
	\item We introduce a general network model that makes it possible to model multiple communication slices running over the same infrastructure in different configurations, and is apt to represent a large variety of scenarios.
	\item We develop a novel \gls{drl}-based strategy to dynamically orchestrate resource allocation to multiple slices.
	The proposed approach is characterized by high flexibility and can be implemented in different network topologies without the need for additional training. 
	\item We show how \textit{transfer learning} can improve the performance of the system, by specializing the strategy learned by the agents in a scalable fashion.
\end{itemize}

\edit{The performance of the proposed strategy is assessed in multiple scenarios, including also a real network topology}, and against a Meta-Heuristic technique and an efficient empirical algorithm.

The remainder of the work is organized as follows.
Sec.~\ref{sec:related} discusses the most relevant works in the considered field.
Sec.~\ref{sec:model} describes the system model used for our analysis.
Sec.~\ref{sec:drl} recalls the fundamentals of \gls{drl} and describes our learning architecture. 
Sec.~\ref{sec:strategy} presents the resource allocation strategies used a benchmark. 
Sec.~\ref{sec:setting_results} describes the simulation scenario and presents the results of our research.
Finally, Sec.~\ref{sec:conclusion} concludes the paper with a recap of the lessons learned and some ideas for future work.

\section{Related Work}\label{sec:related}

Future telecommunication systems will be characterized by the progressive softwarization of network functionalities \edit{and increase of service heterogeneity.
To deal with such a scenario, it is necessary to design new strategies enabling the fully sharing of physical and virtual resources by means of the \gls{ns} paradigm.}
In this respect, the authors of~\cite{halabian2019} analyze a 5G scenario with end-to-end slices contending for the virtual resources offered by data centers, proposing a fully distributed algorithm to maximize system performance. 
In~\cite{leconte2018}, Leconte \emph{et al.} design a \gls{ns} model where multiple traffic flows share network bandwidth and cloud processing units; hence, they implement the Alternating Direction Method of Multipliers~\cite{boyd2011} to determine the best resource allocation scheme. 
\edit{In~\cite{xiao2018} it is adopted a similar approach in a system where} multiple network operators share both licensed and unlicensed spectrum. 
\edit{Besides, the authors of \cite{hu2019} focus on the problem of offloading user tasks to edge computing facilities and design a novel algorithm to optimize resource utilization.}
Finally, Fossati \emph{et al.} propose a framework to generalize multi-resource allocation techniques according to different fairness goals, considering also the critical scenario where resources are not sufficient to satisfy all the slices' demands~\cite{fossati2020}. 

\edit{To address the many challenges related to the \gls{ns} management}, the scientific community has shown great interest in implementing \gls{ml} techniques in such scenarios.
In~\cite{alvizu2017}, the authors exploit \glspl{nn} to predict the traffic evolution in a mobile core-network, thus optimizing the routing and the wavelength assignment according to the \gls{sdn} principles.
\edit{Another example can be found in \cite{hua2020}, where generative adversarial \glspl{nn} are used to minimize the noise in the measurement of \gls{sla} satisfaction.
Instead, the authors of \cite{thantharate2019} design a system based on convolutional \glspl{nn} to associate users with network slices according to the required \gls{qos}.
Finally, in~\cite{Saputra19}, it is implemented a distributed architecture predicting the amount of data that has to be cached in the network edge to address the user demands.
}

\edit{Among all the \gls{ml} techniques that are used for slice orchestration, \gls{drl} is particularly appreciated since its ability to learn complex strategies by trial and error, without the need of labeled data.
In~\cite{huynh2019}, it is defined a novel resource allocation policy, based on Q-Learning~\cite{watkins1992}, that jointly handles the bandwidth, computational and storage requirements of slice users. 
Instead, in~\cite{ayala2019}}, Ayala-Romero \emph{et al.} investigates the orchestration of virtualized radio resources by means of a \gls{drl} framework that encodes traffic data features into resource control decisions.
A similar approach is considered in~\cite{roig2019}, where virtual network functions \edit{are dynamically reconfigured in order to maximize the \gls{qos} of slice users and minimize the overall system cost.}
\edit{Besides, Abiko \emph{et al.} develop a multi-agent architecture to distribute radio resource blocks and prove its adaptability to a variable number of slices~\cite{abiko2020}.}
Finally, the authors of~\cite{alsenwi2020} propose a \gls{drl} \edit{system} to balance the communication requirements of \gls{embb} and \gls{urllc} slices; particularly, an Actor-Critic algorithm is used to schedule \gls{urllc} transmissions without degrading the performance of the \gls{embb} flows. 

Despite the growing interest in this domain, many open questions still need investigation. 
\edit{Most of the aforementioned approaches, indeed, consider that it is always possible to manage the network in a centralized fashion, without addressing the problem of optimize distributed systems where the network status is only partially observable.
For instance, the authors of~\cite{huynh2019} assume that the slice requirements can be always satisfied by the network infrastructure, which is seen as an unique element with an aggregated rate, computational and storage capacity. 
Instead, the work presented in~\cite{roig2019} focuses on the allocation of virtual network functions, considering a standardized system that can be hardly be adapted to different scenarios.
Besides, the authors of~\cite{alsenwi2020} analyze the distribution of the radio frequency blocks in the access network, without taking into account the interaction between the other elements of the network. 
Finally, neither of the above works thoroughly investigate the adaptability of the proposed solutions to different network topologies, which is a key aspect of slice orchestration.
}

\edit{The great heterogeneity of future telecommunication systems requires the implementation of more flexible strategies, which enable the coexistence of multiple services and can promptly adapt to new resource demands. 
A very promising solution is to exploit hierarchical reinforcement learning, which is an approach that has not yet been fully investigated in this context.}
Moreover, the transfer learning paradigm can be used to improve the training of the learning agents~\cite{pan2009}, thus increasing the system adaptability to multiple scenarios.
Our work develops along with these directions, with the final aim of designing a fully scalable \gls{drl} system that can be separated into smaller units, capable of both acting autonomously and cooperating to orchestrate network resources under multiple working conditions.

\section{System Model}\label{sec:model}

\begin{table*}[h!]
	\centering
	\footnotesize
	\caption{Model parameters.}
	\label{tab:model_param}
	\begin{tabular}[c]{@{}cc|cc|cc@{}}
		\toprule
		Parameter & Description & Parameter & Description & Parameter & Description \\
		\midrule
		
		$\phi \in \Phi$ & Information flow & $\epsilon^i_\phi,\epsilon^e_\phi$ & Flow endpoints & $B_l$ & Link rate capacity [bps] \\
		$\sigma \in \Sigma$ & Slice class & $\mathbf{r}_\phi$ & Flow demand vector & $C_n^c$ & Node computational capacity [bps] \\
		$l \in \mathcal{L}$ & Link & $\rho_\phi$ & Resource required by $\phi$ & $C_n^m$ & Node memory capacity [b] \\
		$n \in \mathcal{N}$ & Node & $\hat{\rho}_\phi$ & Resource assigned to $\phi$ & $b_{l, \phi}^i$ & Input flow rate [bps] \\
		$\eta$ & Throughput [bps] & $f_{\sigma}(\cdot)$ & Resource performance function & $b_{l, \phi}^o$ & Output flow rate [bps] \\
		c & Computational power [bps] & $F_{\sigma}(\cdot)$ & Flow performance function & $\tau_n$ & Node routing delay [s] \\
		m & Memory capacity [b] & $\Omega$ & System utility & $D_{l, \phi}$ & Data of $\phi$ queued in $l$ [b] \\
		$\delta$ & Delay [s] & $b_{l, \phi}$ & Bit rate assigned by $l$ to $\phi$ [bps] & $\tau_{l,\phi}^q$ & Queuing delay of $\phi$ in $l$ [s] \\
		$t$ & Discrete time & $c_{n,\phi}$ & Computation assigned by $n$ to $\phi$ [bps] & $\tau_{l,\phi}^\tau$ & Transmission delay of $\phi$ in $l$ [s] \\
		$T$ & Timeslot duration [s] & $m_{n,\phi}$ & Memory assigned by $n$ to $\phi$ [b] & $\tau_{l}^p$ & Link propagation delay [s] \\
		
		\bottomrule
	\end{tabular}	
\end{table*}

In this section, we model a \gls{ns} environment where multiple information flows contend for the same physical and virtual network resources.
We adopt a fluid traffic model, where the traffic through a link is viewed as a fluid stream of data with a certain flow rate.
In particular, we assume that network slices are organized hierarchically, so that more flows can be compound into an aggregate slice, possibly managed by a different tenant. 
The resulting framework is thus very flexible, and can model the interactions among slice tenants, and with slice brokers.
For the reader convenience, we report the main parameters of our model in Tab.~\ref{tab:model_param}.

\subsection{Slice Model}\label{sec:slice}

In our system, we define a \textit{network slice} as an aggregation of information flows with similar behaviors and requirements.
We denote by $\Sigma$ the set including all the different classes of slices, and by $\Phi$ the set including all the information flows.
Given $\sigma \in \Sigma$, we indicate by $\Phi_\sigma$ the set of all information flows belonging to $\sigma$.
Each information flow $\phi \in \Phi$ is characterized by a tuple of parameters, namely:
\begin{itemize}
	\item the \emph{flow endpoints} $\mathcal{E}_\phi=(\epsilon^i_\phi,\epsilon^e_\phi)$, which correspond to the network nodes where the users’ data enter/exit the slice and usually correspond to base stations, edge routers of autonomous systems, or servers;
	\item the \emph{resource demand vector} $\mathbf{r}_\phi=[\eta_\phi, c_\phi, m_\phi, \delta_\phi]$, whose elements are the requirements in terms of throughput ($\eta$), computational power ($c$), memory capacity ($m$), and delay ($\delta$) of the flow;
	\item the \emph{performance function} $F_\phi(\cdot)$, which describes the performance of the considered information flow according to the level the SLA is fulfilled.
\end{itemize}
\edit{We assume that $\mathcal{E}_\phi$, $F_\phi(\cdot)$ and $\delta_\phi$ do not change for the whole duration of flow $\Phi$, while $\eta_\phi$, $c_\phi$ and $m_\phi$ may change in time depending on the dynamic of the data source.}
Note that the throughput is measured in bits per second [b/s], the memory capacity in bits
[b], and the delay in seconds [s]. Finally, we assume the computational requirements are somehow related to data generated by the source. We hence define the computational power as the speed at which data are processed in the computing facilities, and expressed it in bits per second [bps].

In our framework, the time is discretized in timeslots of $T$ seconds, and the information flow parameters can change only slot by slot. 
We write $\mathbf{r}_\phi(t)$ to indicate the resource demand vector of $\phi$ during timeslot $t$, while we write $\hat{\mathbf{r}}_\phi(t) = [\hat{\eta}_\phi(t)$, $\hat{c}_\phi(t)$, $\hat{m}_\phi(t)$, $\hat{\delta}_\phi(t)]$, to indicate the resources assigned to $\phi$ during timeslot $t$.
Note that $\mathbf{r}_\phi(t)$ is determined by the slice class $\sigma$ that $\phi$ belongs to, while $\hat{\mathbf{r}}_\phi(t)$ is determined by the resource allocation strategy. 

As mentioned, we consider two different slice classes: \gls{embb} ($e$) and \gls{urllc} ($u$).
In particular, given $\Phi_e$ (i.e., the set of the \gls{embb} flows) and $\Phi_u$ (i.e., the set of the \gls{urllc} flows), we have $\Phi = \Phi_e \cup \Phi_u$ and $\Phi_e \cap \Phi_u = \emptyset$.
Hence, all the information flows belonging to the same slice $\sigma$ shares the same performance function $F_{\sigma}(\cdot)$, i.e., $\forall \phi \in \Phi_{\sigma}, \, F_{\phi}(\cdot)=F_{\sigma}(\cdot)$. 
In general, $F_\sigma(\cdot)$ depends on both $\mathbf{r}$ and $\hat{\mathbf{r}}$, and returns a value in $[0,1]$, where $1$ means that the \gls{sla} has been completely fulfilled. 
\edit{We assume that $F_\sigma(\cdot)$ is a combination of four functions $f_\sigma(\cdot)$, each of those returns the flow performance for a specif resource $\rho \in \{\delta,\eta,c,m\}$.}
Particularly, $f_{\sigma}(\cdot)$ takes as input $x_\rho$, which is the level of fulfillment of the flow demand, and can take a different shape as long as $f_{\sigma}(x)=1$ $\forall$ $x \geq 1$, i.e., the performance is maximized anytime the allotted resource equals or exceeds the request.

\begin{figure}[h!]
	\centering
	\includegraphics[width=0.9\linewidth]{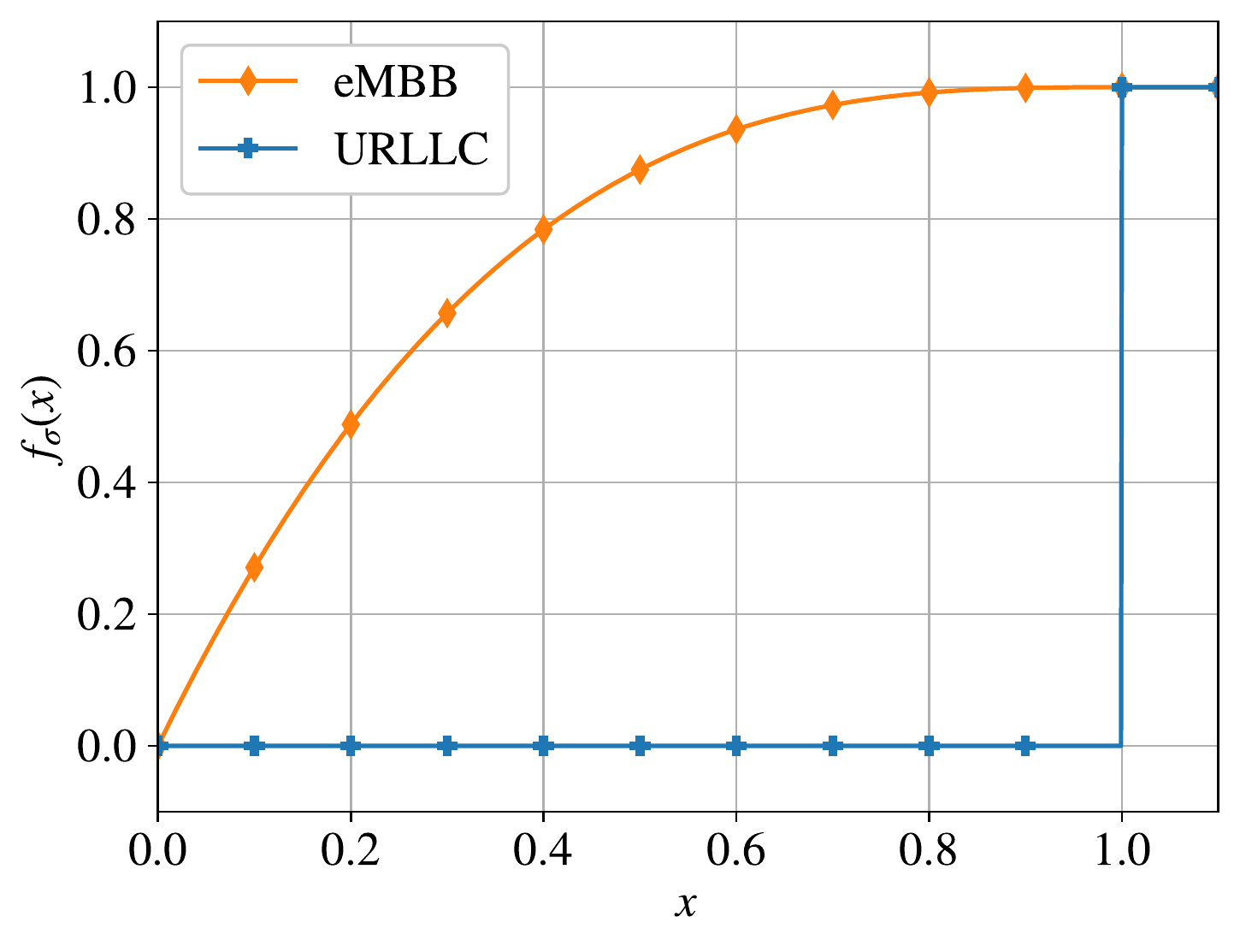}
	\caption{Resource performance function.}
	\label{fig:perf}
\end{figure}

For what concerns the \gls{embb} slice, we assume
\begin{equation}
\label{eq:perf_e}
F_{e}(\mathbf{r}, \hat{\mathbf{r}}) = \alpha_{\delta} f_{e}\left(\frac{\delta}{\hat{\delta}}\right) + \sum_{\rho \in \{\eta, c, m  \}} \alpha_{\rho} f_{e} \left( \frac{\hat{\rho}}{\rho} \right)
\end{equation}
where $\alpha_{\eta}, \alpha_{c}, \alpha_{m},$ and $\alpha_{\delta}$ are non negative and add up to $1$.
Hence, $F_e(\cdot)$ is the weighted sum of $f_e(x_\rho)$, which is a convex function defined as
\begin{equation}
\label{eq:f_e}
f_e(x) = \begin{cases}
\beta_1 x + \beta_2  x^2 + \beta_3 x^3, \quad & x \in [0,1); \\
1, \quad & x \geq 1.
\end{cases}
\end{equation}
Particularly, $\beta_1$, $\beta_2$ and $\beta_3$ are scalar parameters ensuring that $f_e(\cdot)$ is concave and monotonic increasing for $x \in [0,1]$. 
The smooth and concave shape of $f_e(\cdot)$ shown in Fig.~\ref{fig:perf} embodies the flexibility of the \gls{sla} for \gls{embb} services.
\edit{Hence, we assume that the quality of experience of the slice users degrades rather graciously when the \gls{sla} is violated, as in the case of video-streaming applications~\cite{Kimura21}.}

Conversely, \gls{urllc} flows have very strict requirements that, if infringed, cause the sudden degradation of the related services.
\edit{This is intended to represent the fragility of applications such as robotic surgery, which do not tolerate any increase in the communication delay.} 
For this reason, the performance function for this class of services is defined as the product of step functions $f_u(\cdot)$ (shown in Fig.~\ref{fig:perf}):
\begin{equation}
\label{eq:perf_u}
F_u (\mathbf{r}, \hat{\mathbf{r}}) = f_{u}\left(\frac{\delta}{\hat{\delta}}\right) \times \prod_{\rho \in \{\eta, c, m \} } f_{u} \left(\frac{\hat{\rho}}{\rho}\right),
\end{equation}
where 
\begin{equation}
\label{eq:f_u}
f_{u}(x) = \begin{cases}
0, \quad & x \in [0,1); \\
1, \quad & x \geq 1.
\end{cases}
\end{equation}
In this case, $F_u(\cdot)$ drops to zero as soon one single resource requirement is not met.

We remark that our system can be easily extended by defining other slices with different performance functions. 
\edit{Particularly, the slice broker can take advantage of the generality of our system and, for instance, change the composition of the slice set, in order to address the requirements of new tenants in different scenarios.}

Given the functions $F_{\sigma}(\cdot)$ of all slices $\sigma \in \Sigma $, the system utility is obtained as
\begin{equation}
\label{eq:system_perf}
\Omega = \frac{|\Phi_e|}{|\Phi|}  \Omega_e + \frac{|\Phi_u|}{|\Phi|}  \Omega_u, 
\end{equation}
where
\begin{equation}
\label{eq:slice_perf}
\Omega_{\sigma}= \frac{1}{|\Phi_\sigma|} \sum_{\phi \in \Phi_\sigma} F_\sigma(\mathbf{r}_{\phi}, \hat{\mathbf{r}}_{\phi}), \text{ } \sigma \in \Sigma,
\end{equation}
and $|\mathcal{X}|$ represents the cardinality of $\mathcal{X}$. 
We observe that $F_\sigma(\cdot)$ always takes values in $[0,1]$, so that we also have $\Omega \in [0,1]$. 

\edit{Besides, we can define the system utility for a specific type of resource $\rho$ as
\begin{equation}
\label{eq:system_res_perf}
\Omega^\rho = \frac{|\Phi_e|}{|\Phi|}  \Omega_e^\rho + \frac{|\Phi_u|}{|\Phi|}  \Omega_u^\rho, 
\end{equation}
where
\begin{equation}
\label{eq:slice_res_perf}
\Omega_{\sigma}^\rho= \begin{cases}
& \frac{1}{|\Phi_\sigma|} \sum_{\phi \in \Phi_\sigma} f_\sigma\left(\rho / \hat{\rho}\right), \text{ if } \rho = \delta; \\
& \frac{1}{|\Phi_\sigma|} \sum_{\phi \in \Phi_\sigma} f_\sigma\left(\hat{\rho} / \rho\right), \text{ otherwhise};
\end{cases}
\end{equation}
and, as before, $\sigma \in \Sigma$.}

\subsection{Network Model}\label{sec:network}

Our model is based on two different network elements, namely \textit{node} and \textit{link}, as detailed below.
\begin{itemize}
	\item We distinguish two types of \text{nodes}: \textit{access nodes} are located at the network edge and connect users with the rest of the network; 
	\textit{core nodes} are located in the core of the network and forward the aggregated data flows coming from the access nodes.
	Each node $n$ is equipped with a certain amount of computational $C_n^c$ and memory $C_n^m$ resources, which may differ between access and core nodes.
	\item We call \textit{link} any connection $l$ between two different nodes of the network (fronthaul or backhaul).
	This element is provided with a certain bit rate $B_l$ to support the communications between the connected nodes. 
\end{itemize}
From now on, we denote by $\mathcal{N}$ and $\mathcal{L}$ the set of network nodes and links, respectively. 
Particularly, $\mathcal{N}$ can be partitioned into $\mathcal{N}_a$, which includes the access nodes, and $\mathcal{N}_c$, which includes the core nodes.

In our model, each information flow $\phi \in \Phi$ is initialized between two access nodes and passes through a certain number of core nodes and links.
Let $\Phi_{l}$ and $\Phi_{n}$ be the set of information flows that cross link $l \in \mathcal{L}$ and node $n \in \mathcal{N}$, respectively.
We assume that each flow $\phi$ always goes through the same network elements from $\epsilon_{\phi}^i$ to $\epsilon_{\phi}^e$, which implies that $\Phi_{l}$ and $\Phi_{n}$ do not change in time $\forall$ $l \in \mathcal{L}$, $n \in \mathcal{N}$.

\edit{We observe that, in general, slices can be activated and deactivated on-demand, thus varying the number of flows in the network. 
Nonetheless, once established, the path of an information flow is generally maintained for the whole duration of the connection, unless some links become unavailable or the communication end points change.
In this case, the resource allocation framework will react as if the flow was interrupted an a new one was started along the new path.
We assume that such events are rare and do not impact significantly on the performance of the proposed scheme.}

Given a link $l \in \mathcal{L}$, each flow $\phi \in \Phi_{l}$ gets assigned a portion $b_{l,\phi}$ of the link bit rate $B_l$.
Similarly, each node $n \in \mathcal{N}$ assigns to $\phi$ an amount $c_{n,\phi}$ and $m_{n,\phi}$ of its computational and storage resources. 
Consequently, any resource allocation pattern must comply with the following feasibility conditions:
\begin{align}
& \sum_{\phi \in \Phi_l} b_{l,\phi} \leq B_l, & \forall \text{ } l \in \mathcal{L}; \label{eq:condition_b}\\
& \sum_{\phi \in \Phi_n} \rho_{n,\phi} \leq C_n^\rho,
& \forall \text{ } n \in \mathcal{N}, \rho \in \{c, m\}. \label{eq:condition_c}
\end{align}

Given a certain allocation of network resources, we want to compute $\hat{\mathbf{r}}_\phi$, $\forall$ $\phi \in \Phi$.
We denote by $\mathcal{N}_\phi$ and $\mathcal{L}_\phi$ the ordered set of network nodes and links crossed by $\phi$, respectively.
In particular, the first and the last element of $\mathcal{N}_\phi$ constitute the flow endpoints.
We assume that the computational and memory requests of a flow $\phi \in \Phi$ can be distributed among all the nodes in $\mathcal{N}_\phi$, so that
\begin{gather}\label{eq:const_8}
\hat{\rho}_\phi = \sum_{n \in \mathcal{N}_\phi} \rho_{n,\phi}, \quad \forall \text{ } \rho \in \{c, m\}. 
\end{gather}

To determine the throughput $\hat \eta_\phi$, instead, we need to consider the output flow rate $b_{l,\phi}^o(t)$ at time $t$ from each link $\l\in\mathcal{L}$. Indeed, the throughput corresponds to the output rate from the last link $\lambda$ along the path: 
\begin{equation}\label{eq:const_9}
\hat{\eta}_\phi(t) = b_{\lambda,\phi}^o(t).
\end{equation}
In turn, $b_{l,\phi}^{o}(t)$ depends on both the bit rate $b_{l,\phi}(t)$ assigned to $\phi$ by $l$, the input flow rate $b_{l,\phi}^{i}(t)$ from the upstream link, and the amount of data of $\phi$ queued at node $n$ at the end of the previous slot, which is denoted by $D_{l,\phi}(t-1)$.
The input flow rate $b_{l,\phi}^{i}(t)$ is given by
\begin{equation}
b_{l,\phi}^{i}(t) = \begin{cases}\label{eq:const_10}
b_{\ell,\phi}^{o}(t), &\text{ if } \ell \text{ is the upstream link of $l$ in } \mathcal{L}_\phi;\\
\eta_\phi(t), &\text{ if } l \text{ is the first link in } \mathcal{L}_\phi.
\end{cases}
\end{equation}

The output flow rate $b_{l,\phi}^{o}(t)$ is then given by the minimum between the allocated rate $b_{l,\phi}(t)$ and the sum of the incoming and queued traffic, i.e.,
\begin{equation}\label{eq:const_11}
b_{l,\phi}^{o}(t) = \min\left\{b_{l,\phi}(t), \frac{D_{l, \phi}(t-1)}{T} + b_{l,\phi}^{i}(t) \right\}.
\end{equation}
The variable $D_{l, \phi}(t)$ is set to $0$ for any time $t$ before the initialization of the flow, and then it is updated as 
\begin{equation}\label{eq:const_12}
D_{l, \phi}(t) = 
\max\left\{0, D_{l, \phi}(t-1) + T (b_{l,\phi}^i(t) - b_{l,\phi}(t)) \right\}.
\end{equation}
Note that the value of $D_{l, \phi}$ increases as the assigned rate $b_{l,\phi}$ is lower than the input rate $b_{l,\phi}^i$.

For what concerns the delay experienced by $\phi$, we have
\begin{equation}\label{eq:const_13}
\hat{\delta}_\phi(t) = \sum_{n \in \mathcal{N}_\phi} \tau_n + \sum_{l \in \mathcal{L}_\phi} \tau_{l, \phi}(t),
\end{equation}
where $\tau_n$ is a positive value representing the delay due to routing operations at node $n$, and it is assumed constant over time.
Instead, $\tau_{l,\phi}(t)$ is computed as
\begin{equation}\label{eq:const_14}
\tau_{l,\phi}(t) = \tau^{q}_{l,\phi}(t) + \tau^{\tau}_{l,\phi}(t)+ \tau^{p}_{l},
\end{equation}
where $\tau^{q}_{l,\phi}$, $\tau^{\tau}_{l,\phi}$, and $\tau^{p}_{l}$ represent the queuing, transmission, and propagation delays of $\phi$ through link $l$, respectively. 
In particular, $\tau^{q}_{l,\phi}(t)$ is the average queuing time of a bit in $l$ during $t$, and is given by (see the Appendix)
\begin{equation}\label{eq:const_15}
\tau^{q}_{l,\phi}(t) = \frac{2 D_{l, \phi}(t - 1) - T (b_{l, \phi}^{o}(t) - b_{l, \phi}^{i}(t))}{2 b_{l, \phi}(t)}.
\end{equation}
Instead, $\tau^{\tau}_{l,\phi}(t)$ is the reciprocal of $b_{l,\phi}^{o}(t)$, i.e.,
\begin{equation}\label{eq:const_16}
\tau^t_{l,\phi}(t) = \frac{1}{b_{l,\phi}^{o}(t)}.
\end{equation}
Finally, $\tau^{p}_{l}$ is a positive and constant value that depends on the physical characteristic of the communication link. 
We highlight that, despite we consider a discrete time-frame, $\hat{\delta}_\phi$ is a continuous value. 

Our aim is to determine the best resource allocation to maximize the system utility as given in \eqref{eq:system_perf}.
Mathematically, we want to determine $\hat{\mathbf{r}}_\phi$, $b_{l,\phi}$, $c_{n,\phi}$, $m_{n,\phi}$, $b^i_{l,\phi}$,  $b^o_{l,\phi}$, $\tau^q_{l,\phi}$, and $\tau^{\tau}_{l,\phi}$, $\forall$ $\phi \in \Phi$, $n \in \mathcal{N}$, $l \in \mathcal{L}$ that maximize $\Omega$, under the constraints given in \eqref{eq:condition_b}-\eqref{eq:const_16}.

The many constraints and the non-convexity of $\Omega$ make the problem very complex to solve.
\edit{In particular, the optimal solution can be determined only if a central controller is provided with all the system variables at any timeslot. Then, conventional optimization tools or meta heuristic techniques can be used to identify the best resource allocations scheme. 
However, the first may converge on local maxima, while the latter can be unable to find a solution within a reasonable time frame.
If the resource demands of the information flows evolve quickly in time, a valuable approach is to implement distributed control algorithms that can promptly take new actions, albeit with a partial view on the overall network.
In particular, the DRL paradigm is particularly suitable for this problem since it can provide high-performance solutions to carry our complex control tasks also when the environment is partially observable.}

\section{Learning Strategy}\label{sec:drl}

In order to efficiently orchestrate communication resources in a \gls{ns} scenario, we develop a distributed architecture based on multiple learning units, named \textit{local controllers}, that collaborate to maximize the overall system utility given by \eqref{eq:system_perf}.
In the rest of the section, we will recall the main principles of \gls{drl}, and present the framework used to train the learning agents of our architecture.

\subsection{Deep Reinforcement Learning}

The \gls{rl} paradigm is one of the main branches of \gls{ml}.
Particularly, \gls{rl} does not solely aim at solving classification or regression problems, but it enables the development of complex strategies to maximize the long-term performance of a target system~\cite{sutton1998introduction}.
Moreover, \gls{rl} algorithms do not need to have labeled data to carry out the training phase, but they interact with a learning environment where agent actions are associated to specific rewards. 

In a \gls{rl} scenario, the target system is modeled as a \gls{mdp}, which is a powerful mathematical tool used to represent decision-making problems~\cite{puterman2014}. 
This framework requires to define a state space $\mathcal{S}$ of the environment, an action space $\mathcal{A}$ of the learning agent, and a reward function $r: \mathcal{S} \times \mathcal{A} \rightarrow \mathbb{R}$. 
During any timeslot $t$, the learning agent observes the system state $s_t \in \mathcal{S}$, performs an action $a_t \in \mathcal{A}$ and receives a reward $r_t \in \mathbb{R}$.
Hence, the future state $s_{t+1}$ depends uniquely on the previous state $s_t$ and the agent decision $a_t$. 

The agent chooses new action according to a policy $\pi: \mathcal{S} \times \mathcal{A} \rightarrow [0,1]$, where $\pi(s, a)$ is the probability to take the action $a$ when the state $s$ is observed.
The aim of any \gls{rl} algorithm is to determine a policy that maximizes the system long-term reward, which is
\begin{equation}
	R = \sum_{t=0}^{\infty} \lambda^t r_t,
\end{equation}
where $\lambda \in [0,1]$ is the so-called \textit{discount factor}.
Particularly, if $\lambda \rightarrow 0$, the algorithm favors the actions that can acquire high reward in a short time; instead, if $\lambda \rightarrow 1$, the algorithm aims at determining the actions that bring more benefit in the future. 

Given a policy $\pi$, a \gls{rl} algorithm associates each possible state $s$ with a value $V_\pi(s)$.
The function $V_\pi(\cdot)$ is called state value function and represents the expected cumulative reward that is achieved following the actions of $\pi$ from state $s$.
In other words, we have
\begin{equation}
V_\pi(s) = E\left[ R | s, \pi \right].
\end{equation}
As the agent explores the learning environment, the values of $V_\pi(s)$ and the policy itself are updated. 
In this perspective, the optimal policy $\pi^*$ provides the maximum value of $V_\pi(s)$ $\forall$ $s \in \mathcal{S}$, i.e.,   
\begin{equation}
	V_{\pi^*}(s) = \max_\pi V_\pi(s), \quad \forall \text{ } s \in \mathcal{S}.
\end{equation}

If the state and action spaces get too complex, conventional \gls{rl} approaches fail to determine the optimal policy because of the \textit{curse of dimensionality}~\cite{indyk1998}.
To address such a problem, the \gls{drl} paradigm allows to approximate the state value function and the optimal policy by deep \glspl{nn}.
In particular, \gls{drl} algorithms are capable of handling continuous state and action spaces, which means that $|\mathcal{S} \times \mathcal{A}| \rightarrow \infty$. 

\subsection{Learning Architecture}
\label{sec:drl_arch}

In this work, we adopt an \gls{ac} approach~\cite{konda2000}, which involves the learning of the optimal policy by two different units.
The first is named \textit{actor} and approximates the optimal policy $\pi_\theta$, parameterized by $\theta$; the latter is named \textit{critic} and approximates the value function $V_{\gamma}$, parameterized by $\gamma$. 
Hence, the actor is trained to compute the action $a_t$ that the policy $\pi_\theta$ takes in a state $s_t$; the critic is trained to compute the expected long term reward that is obtained following the policy $\pi_{\theta}$ from $s_t$.

To carry out the system training, we exploit the \gls{a2c} algorithm, which has shown to provide stable \gls{drl} solutions in very complex scenarios~\cite{mnih2016async}. 
The critic is trained to minimize the function $L_c(s_t, a_t) = A(s_t, a_t)^2$, where
\begin{equation}
\label{eq:adv}
A(s_t, a_t) = r_t + \lambda V_\gamma(s_{t+1}) - V_\gamma(s_t).
\end{equation}
Particularly, the function $A: \mathcal{S} \times \mathcal{A} \rightarrow \mathbf{R}$ is called \textit{advantage} and returns the reward gain obtained by choosing action $a_t$ in state $s_t$.
Conversely, the actor is trained to minimize the function 
\begin{equation}
\label{eq:actor}
L_a(s_t, a_t) = - \nabla_\theta \text{log} \pi_\theta(s_t, a_t) A(s_t, a_t)^2 - \kappa H(\pi_\theta),
\end{equation}
where $\nabla_\theta$ is the gradient with respect to $\theta$, $\pi_\theta(s_t, a_t)$ is the probability of taking action $a_t$ in state $s_t$, $H(\pi_\theta)$ is the entropy of $\pi_\theta$, and $\kappa$ is a scalar value.
As suggested in~\cite{mnih2016async}, the actor loss function depends linearly on the policy entropy $H(\pi_\theta)$.
Particularly, we can promote the exploration of the action space by increasing $\kappa$ since, in such a case, the actor gain benefits to take random actions. 
Instead, as $\kappa \rightarrow 0$, the actor will choose actions that are expected to bring the highest reward according to the current experience.

\edit{We highlight that A2C enables to consider continuous action spaces: this is not possible with traditional Reinforcement learning algorithms (e.g., Q-Learning) that, instead, can only take actions from discrete sets. 
Besides, the A2C algorithm supports an online training phase and, consequently, allows the agents to continuously refine the target policy while it is being used in the real system. 
Therefore, the \textit{slice broker} does not need to train a new system from scratch every time the network conditions change since the policy will dynamically adapt to the new conditions as time passes by.}

\begin{figure}[h!]
	\centering
	\includegraphics[width=.9\linewidth]{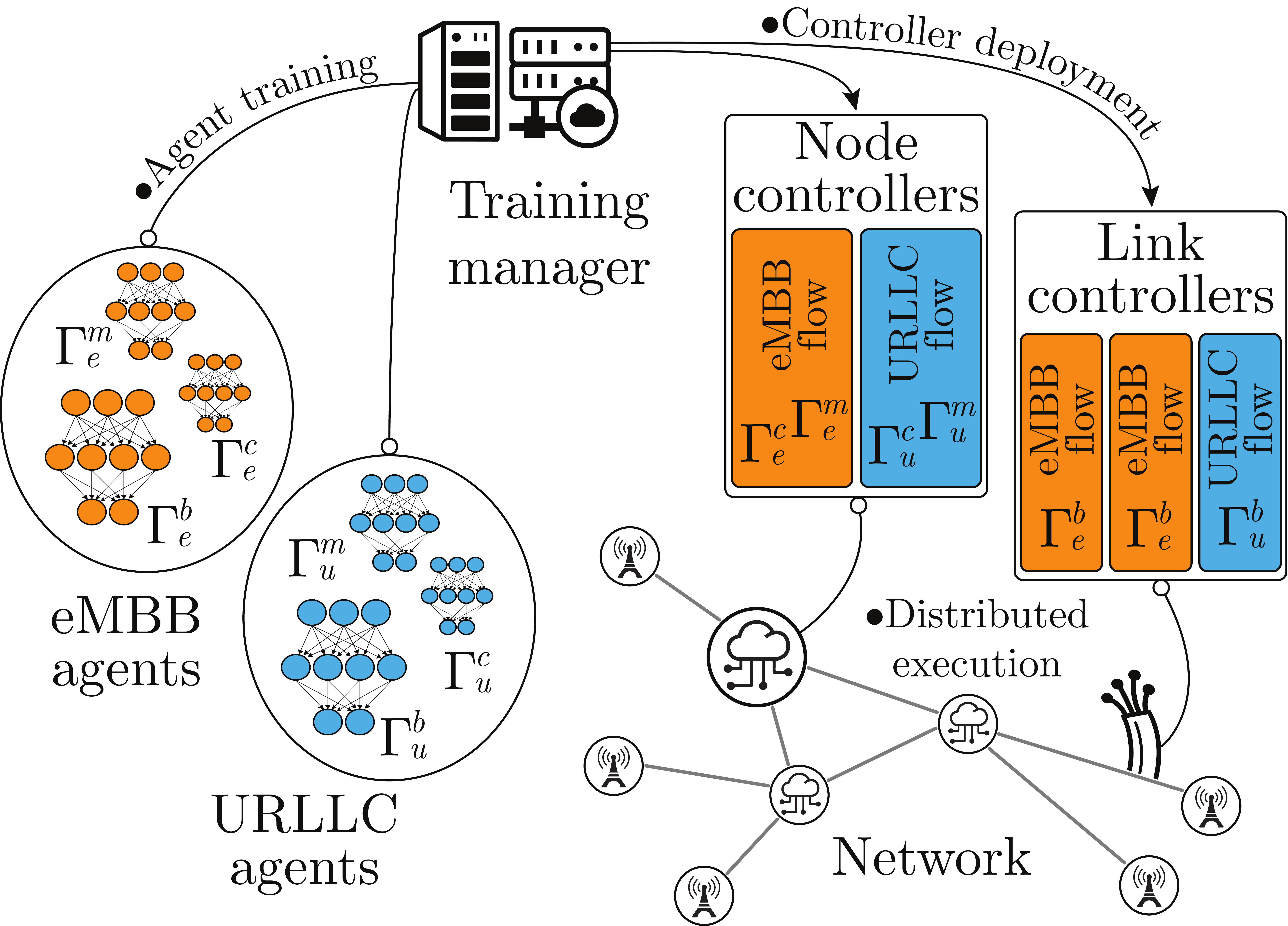}
	\caption{Learning Architecture.}
	\label{fig:architecture}
\end{figure}

Our learning architecture (shown in Fig.~\ref{fig:architecture}) provides a different controller for each information flow and network element.
From an operational point of view, the total number of local controllers depends on the network topology and the cardinality of $\Phi$.
Practically, the local controllers are all replicas of $3 \times |\Sigma|$ learning agents. 
During the training phase, we design a tuples of agents ($\Gamma_\sigma^b$, $\Gamma_\sigma^c$, $\Gamma_\sigma^m$) for each different slice $\sigma \in \Sigma$:
the agent $\Gamma_\sigma^b$ is trained to orchestrate the bit rate of each information flow $\phi \in \Phi_{\sigma}$ in each link $l \in \mathcal{L}$;
instead, $\Gamma_\sigma^c$ and $\Gamma_\sigma^m$ are trained to orchestrate the computation and memory resources of each information flow $\phi \in \Phi_{\sigma}$ in each node $n \in \mathcal{N}$.
\edit{Practically, the training phase is performed by a central entity, named \textit{training manager}, which collect the system information and update the learning architecture accordingly.}
Then, copies of $\Gamma_\sigma^b$, $\Gamma_\sigma^c$, and $\Gamma_\sigma^m$ will be instantiated in each network element crossed by any flow $\phi\in \Phi_\sigma$.

According to the \gls{a2c} algorithm, each learning agent is composed by two units, i.e., the actor and the critic, which are implemented by means of \glspl{nn}.
Particularly, we consider an architecture with two hidden layers and the \gls{relu} as activation function~\cite{zhang2018efficient}.
The output of the actor is the amount $\rho^*$ of resources that the local controller demands, while that for the critic is the expected future reward.
The size of the \gls{nn} input varies according to the type of resources that has to be managed, as explained in the next subsection.

\subsection{Observations and Actions}

In our system, each local controller has full knowledge of the element where it is installed, while it has a limited view on the network, which implies that the system state is only partially observable~\cite{jaakkola1995}. 
Let us consider a local controller managing the rate resources of a flow $\phi$ in a link $l$. 
At the beginning of each timeslot, such a controller is provided with two vectors representing the status of the information flow and the network element that is associated with.

The first vector gives the state of $\phi$ at the beginning of timeslot $t$:
\edit{
\begin{equation}\label{eq:slice_status}
\begin{aligned}
\mathbf{s}_\phi (t) = [ \mathbf{r}_{\phi}(t), \hat{\mathbf{r}}_{\phi}(t-1)],
\end{aligned}
\end{equation}}
where $\mathbf{r}_{\phi}(t)$ and $\hat{\mathbf{r}}_{\phi}(t-1)$, defined in Sec.~\ref{sec:model}, are the resources requested and granted by $\phi$ at the beginning of timeslot $t$ and $t-1$, respectively.
\edit{We observe that $\hat{\mathbf{r}}_{\phi}(t-1)$ can be computed only knowing the aggregate amount of network resources assigned to $\phi$ by the network elements of its routing path.  
Therefore, $\mathbf{s}_\phi(t)$ needs to be shared among all the controllers assigned to $\phi$ at the beginning of each timeslot $t$. 
However, the size of $\mathbf{s}_\phi(t)$ is negligible with respect to the rate requirements of the slices, and, therefore, can be transmitted within the user data plane of $\phi$, without degrading the performance of our system.}

The second vector provides the state of the rate resources of $\phi$ in $l$ at the beginning of timeslot $t$:
\edit{
\begin{equation}
\label{eq:link_status}
\begin{aligned}
\mathbf{s}_{l, \phi}(t) = [ & B_l, \tau_{l,\phi}(t-1), D_{l,\phi}(t-1), b_{l, \phi}^*(t-1), \\ & b_{l, \phi}(t-1),  b_{l}^{e}(t-1), b_{l}^{u}(t-1)],
\end{aligned}
\end{equation}}
where $b_{l}^{\sigma}(t-1)$ is the aggregate rate demanded in $l$ by all the flows of class $\sigma$ during timeslot $t-1$, while the other parameters were defined in Sec.~\ref{sec:model}.
\edit{We highlight that this information is provided by the considered link $l$ and, consequently, the knowing of $\mathbf{s}_{l, \phi}(t)$ does not require any additional communication within the network.}
Hence, at the beginning of timeslot $t$, the link controller takes $\mathbf{s}_\phi (t)$ and $\mathbf{s}_{l, \phi}(t) $ as input and returns $b_{l, \phi}^*(t)$, \edit{which is the bit rate demanded by $\phi$ in $l$ during $t$.}

\edit{When considering a controller associated to a node $n$ and a flow $\phi$ we use the same approach and, depending on the resource $\rho \in \{c, m\}$ we want to allocate, we substitute \eqref{eq:link_status} with $s^c_{n, \phi}(t)$ or $s^m_{n, \phi}(t)$, which are the states of the computation and memory resources assigned to $\phi$ by node $n$.
In particular, $s^c_{n, \phi}(t)$ or $s^m_{n, \phi}(t)$ are defined as}
\edit{
\begin{equation}
\begin{aligned}\label{eq:node_status}
\mathbf{s}^\rho_{n, \phi}(t) = [ & C_n^\rho, \rho_{n, \phi}^*(t-1), \rho_{n, \phi}(t-1), \\ & \rho_{n}^{e}(t-1), \rho_{n}^{u}(t-1) ],
\end{aligned}\\
\end{equation}}
where $\rho_{n}^{\sigma}(t-1)$ is the aggregate amount of resource $\rho$ demanded in node $n$ by all the flows of class $\sigma$ during timeslot $t-1$.
As before, the node controller takes $\mathbf{s}_\phi (t)$ and $\mathbf{s}^\rho_{n, \phi}(t)$ as input and returns $\rho_{n,\phi}^*(t)$, with $\rho \in \{c, m\}$, \edit{which is the amount of computational (or memory) capacity demanded by $\phi$ in $n$ during $t$.}

\subsection{Reward Function}

\edit{In accordance to the RL paradigm, we need to define a reward function $r(\cdot)$ that represents the benefit generated by each possible state-action pair of the policy.
In particular, to maximize the overall utility, each local controller should demand enough resources to maximize the performance of the flow $\phi$ it is in charge of, without subtracting too many resources to the other flows.
In our system, given a controller associated to a link $l \in \mathcal{L}$ and a flow $\phi \in \Phi_l$, the reward at time $t$ is given by
\begin{equation}
\begin{aligned}
\label{eq:reward_l}
r_{l, \phi}(t) = & \gamma_0 \left( f_{\sigma_\phi} \left( \frac{\eta_\phi(t) }{\hat{\eta}_\phi(t)} \right) + f_{\sigma_\phi} \left( \frac{\hat{\delta}_\phi(t) }{\delta_\phi(t)} \right) \right) + \\
& \frac{\gamma_1}{|\Phi_l|} \sum_{\psi \in \Phi_l } \left( f_{\sigma_\psi} \left( \frac{\eta_\psi(t) }{\hat{\eta}_\psi(t)} \right) + f_{\sigma_\psi} \left( \frac{\hat{\delta}_\psi(t) }{\delta_\psi(t)} \right) \right),
\end{aligned}
\end{equation}
where $\sigma_\phi$ is the slice class that $\phi$ belongs to, while $\gamma_0$ and $\gamma_1$ are positive scalar values.
In particular, $\gamma_0$ weights the throughput and delay performance of flow $\phi$, while $\gamma_1$ weights the average throughput and delay performance of all the other flows crossing link $l$.}

\edit{
Similarly, a controller assigned to a node $n$ and a flow $\phi$ is rewarded according to
\begin{equation}
\label{eq:reward_m}
r_{n, \phi}^\rho(t) = \gamma_0 f_{\sigma_\phi} \left( \frac{\rho_\phi(t) }{\hat{\rho}_\phi(t)} \right) + \frac{\gamma_1}{ |\Phi_n|}\sum_{\psi \in \Phi_n } f_{\sigma_\psi} \left( \frac{\rho_\psi(t) }{\hat{\rho}_\psi(t)} \right),
\end{equation}
where $\rho \in \{c, m\}$, while the scalar values $\gamma_0$ and $\gamma_1$ have the same role as before.}

\edit{Therefore, the reward function consists of the weighted sum of two terms: the first reflects the performance of the flow targeted by the agent, while the second represents the aggregate performance of all flows that share that network element.
With such a mixed reward function, the agent will hence attempt to improve the quality of the targeted flow, but without unduly penalizing other flows.}

\section{Benchmark Strategies}\label{sec:strategy}
In this section, we describe the resource allocation strategies that we use as benchmark for our model. 
The first is an empirical algorithm, which tries to fairly allocate communication and processing resources in each network allocation.
The latter is based on meta-heuristic optimization and performs a static allocation of network resources.

\subsection{Empirical Strategy}

Similarly to our approach, the empirical strategy implements a distributed resource allocation scheme.
At the beginning of timeslot $t$, each flow $\phi$ crossing link $l$ demands a bit rate sufficient to both satisfy the current throughput requirement and transmit any buffered data, i.e.,
\begin{equation}
b^*_{l, \phi}(t) = \eta_\phi(t) + \frac{D_{l, \phi}(t-1)}{T}. 
\end{equation}

For what concerns the computation and memory allocation, each flow $\phi$ distributes its requests among the nodes along its path, proportionally to their capacity.
More specifically, the amount of resources required to node $n$ is equal to
\begin{equation}
	\rho^*_{n, \phi}(t) = \chi_{n, \phi} \rho_\phi(t),
\end{equation}
where $\rho \in \{c, m\}$, and $\chi_{n, \phi}$ is computed as
\begin{equation}
	\chi_{n, \phi} = \frac{C_n^\rho} {\sum_{k \in \mathcal{N}_\phi} C_k^\rho }.
\end{equation}

We highlight that, to compute $\chi_{n, \phi}$, it is necessary to known the computation and storage capacities of each node $n \in \mathcal{N}_\phi$.
Such an information is not provided to the local controllers of the \gls{drl} strategy. 
Hence, the empirical strategy has an advantage with respect to our learning framework since it uses information that is generally not available in a fully distributed approach.

We observe that, either using the empirical or the \gls{drl} strategy, network elements may be not able to satisfy all the requests they receive.
Particularly, the total aggregated amount of resources that is demanded to a link $l$ (or a node $n$) may exceed its overall capacity.
Hence, we need to map the demanded resources $b_{l, \phi}^*$, $c_{n, \phi}^*$, $m_{n, \phi}^*$ to the assigned resources $b_{l, \phi}$, $c_{n, \phi}$, $m_{n, \phi}$, $\forall$ $l \in \mathcal{L}$, $n \in \mathcal{N}$, ensuring that the feasibility constraints \eqref{eq:condition_b} and \eqref{eq:condition_c} are always satisfied.

\edit{Let us consider a link $l$ $\in \mathcal{L}$ during a timeslot $t$.
If the total amount of resources demanded at the link is lower than $B_l$, the feasibility constraints are already met: consequently, we can set $b_{l, \phi}(t) = b_{l, \phi}^*(t)$.
In the other case, the rate $b_{l, \phi}(t)$ assigned by the link to each flow $\phi \in \Phi_l$ is proportional to the flow demand $b_{l, \phi}^*(t)$:}
\begin{equation}
b_{l, \phi}(t) = b_{l, \phi}^*(t)  \min \left\{ 1, \frac{B_l}{\sum_{\psi \in \Phi_l} b_{l, \psi}^*(t)}\right\}.
\end{equation}
Using the same principle for the computational and memory resources, we can write
\begin{equation}
c_{n, \phi}(t) =  c_{n, \phi}^*(t)  \min \left\{ 1, \frac{C_n^c}{\sum_{\psi \in \Phi_n} c_{n, \psi}^*(t)} \right\}, 
\end{equation}
and
\begin{equation}
m_{n, \phi}(t) =  m_{n, \phi}^*(t)  \min \left\{ 1, \frac{C_n^m}{\sum_{\psi \in \Phi_n} m_{n, \psi}^*(t)} \right\}.
\end{equation}

\subsection{Static Strategy}

Meta-heuristic techniques have been shown to determine the optimal solution of highly complex optimization problems with non-convex constraints~\cite{yang2010}. 
In our system, a \gls{ga} may be used to approximate the best resource allocation pattern, thus outperforming both the \gls{drl} and empirical strategies. 
However, meta-heuristic algorithms are based on a randomized search of the target solution and require an extremely long calculation time, which makes it unfeasible to execute them at each timeslot.
At the same time, it is reasonable to exploit meta-heuristic techniques if the resource requirements do not vary in time.

In the static strategy we consider in this work, the network resources are statically divided among the different slices.
Practically, each traffic flow $\phi$ is assumed to have fixed requirements, corresponding to the average amount of resources it demands, i.e., $\mathbf{r}_\phi^a$.
Hence, a \gls{ga} is used to determine the optimal resource allocation pattern under such conditions~\cite{whitley1994}.

We observe that, using the static strategy, the values of $b_{l,\phi}, c_{n,\phi}, m_{n,\phi}$ $\forall$ $\phi \in \Phi$, $l \in \mathcal{L}$, $n \in \mathcal{N}$ are maintained fixed. 
In other words, the variability of information flows is not taken into account, and the performance of each flow is expected to deteriorate as soon as its requirements exceed the average values. 
This is a big issue, especially for the \gls{urllc} slice, whose performance function suddenly drops if any resource requirements is not satisfied. 

\begin{figure}[t!]
	\centering
	\includegraphics[width=.9\linewidth]{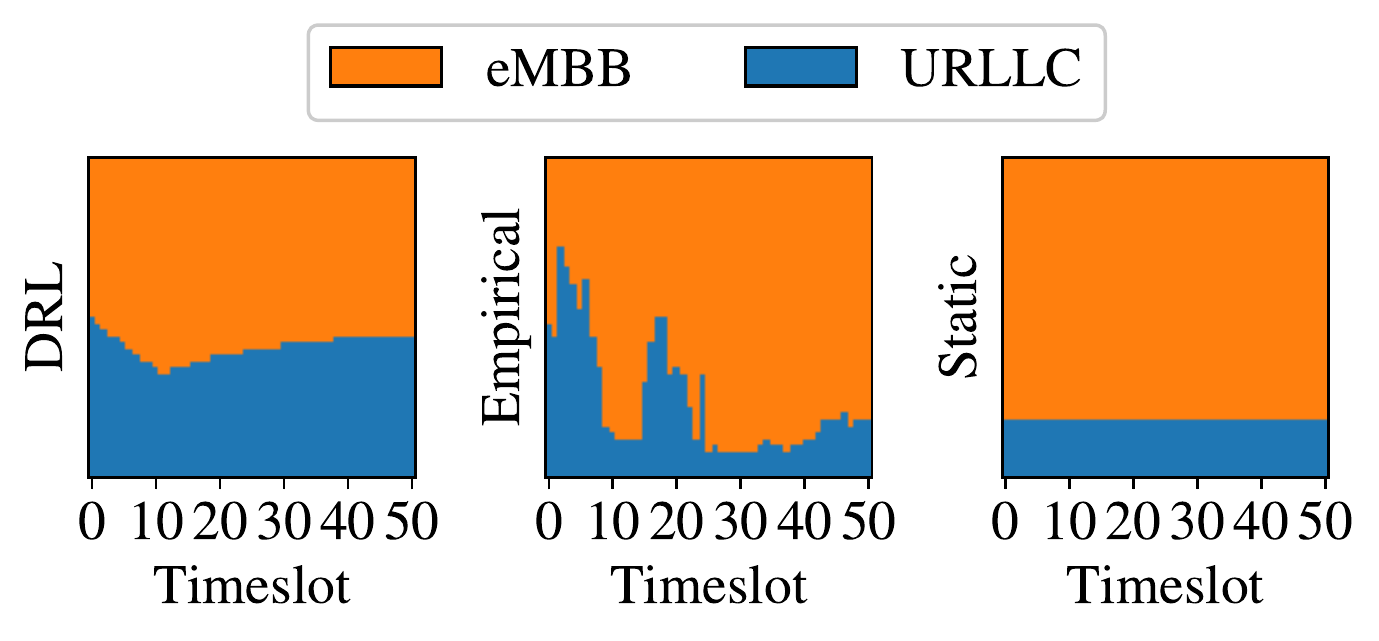}
	\caption{Link resource allocation.}
	\label{fig:link_resource}
\end{figure}

\edit{To better highlight the characteristics of the benchmark strategies, we analyze the bit rate allocation in a network link during a period of $50$ timeslots.
In Fig.~\ref{fig:link_resource}, we represent the share of link resources assigned to the \gls{embb} and \gls{urllc} services, using the different strategies. 
In particular, we consider a scenario where the capacity is lower than the aggregate resource demand, which means that all the strategies fully exploit the link rate. 
As expected, \gls{ga} distributes network resources in a static fashion: the bitrate assigned to the different slices does not vary in time. 
Instead, the empirical algorithm is able to adapt to the slice requirements and, hence, the bitrate distribution changes at each timeslot. 
Also the \gls{drl} strategy follows a dynamic trend but it assigns a larger amount of resources to the \gls{urllc} services than the benchmarks.
}

\section{Simulation Setting and Results}\label{sec:setting_results}

In this section, we first describe the scenarios where our algorithms are tested as well as the setting of our simulations.
Then, we investigate the performance of our \gls{drl} architecture against the benchmark strategies and under different working conditions.
Finally, we show how the transfer learning paradigm can be used to further improve the performance of our system. 

\subsection{Setting}

\begin{figure}[t!]
	\centering
	\begin{subfigure}{0.48\linewidth}
		\centering
		\includegraphics[width=\linewidth]{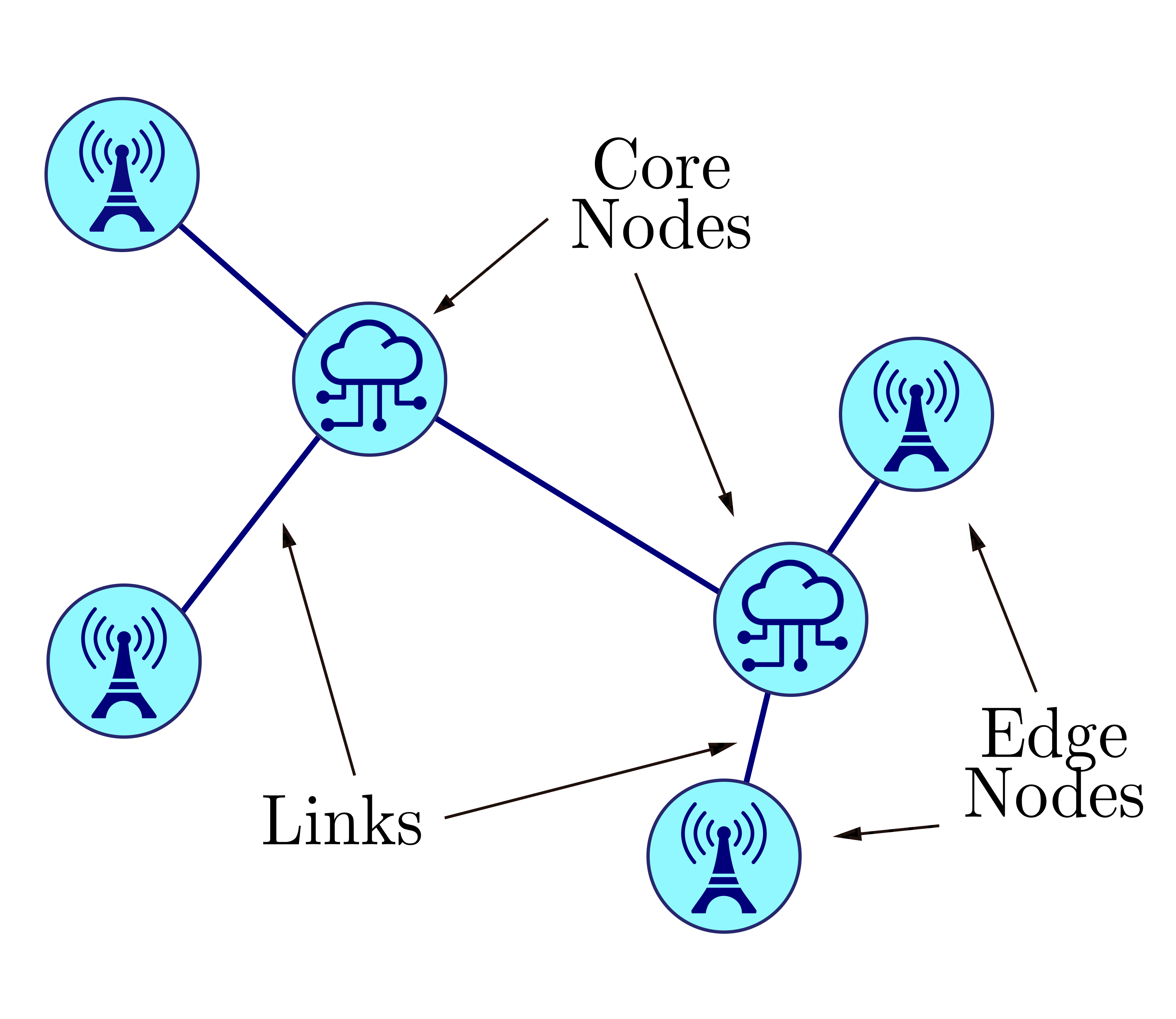}
		\caption{Dumbbell Network (D).}
		\label{fig:sn}
	\end{subfigure}%
	\begin{subfigure}{0.48\linewidth}
		\centering
		\includegraphics[width=\linewidth]{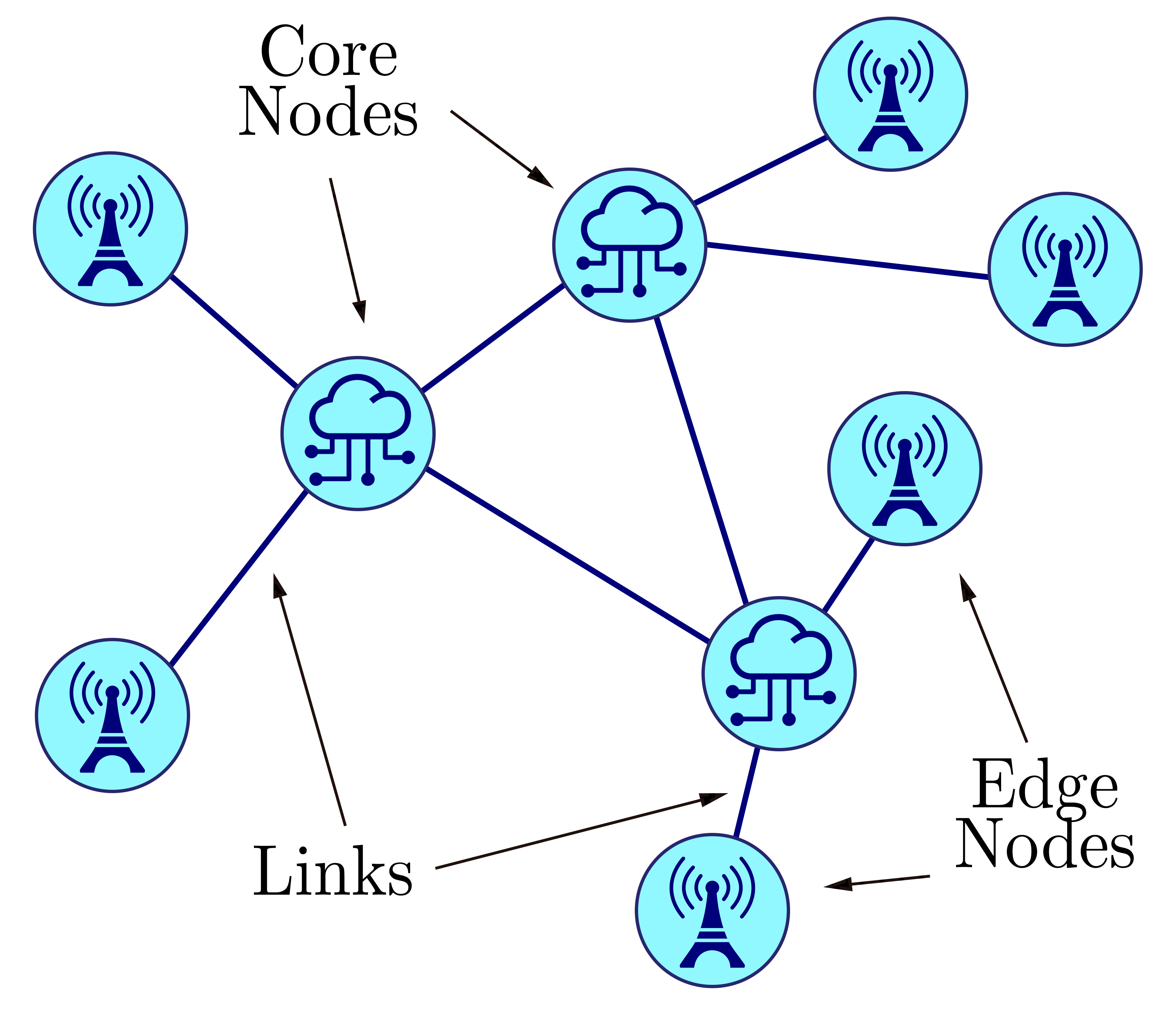}
		\caption{Triangle Network (T).}
		\label{fig:mn}
	\end{subfigure}
	\begin{subfigure}{0.48\linewidth}
		\centering
		\includegraphics[width=\linewidth]{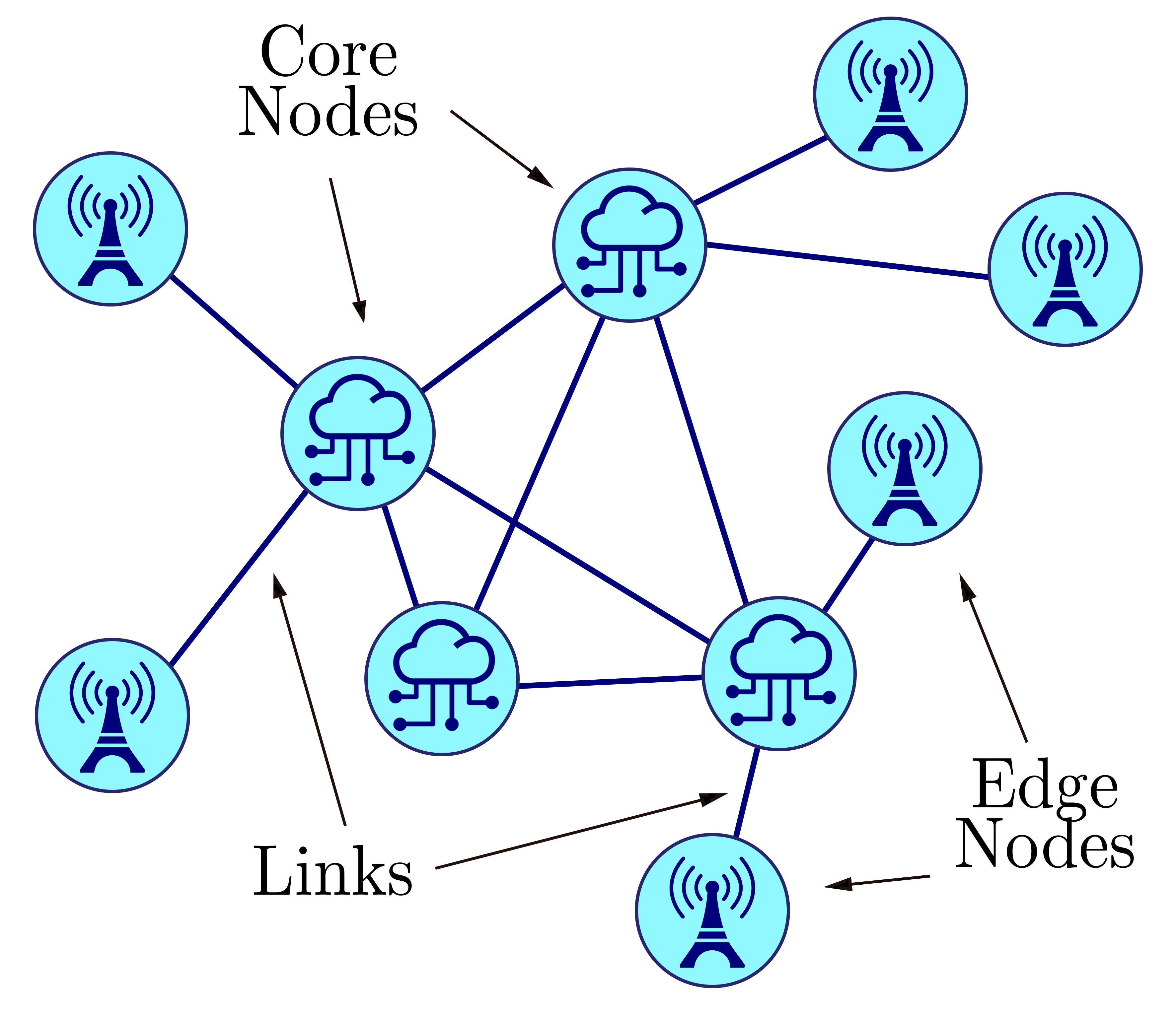}
		\caption{Pyramid Network (P).}
		\label{fig:mn}
	\end{subfigure}
	\caption{Network topologies.}
	\label{fig:networks}
\end{figure}

\edit{We consider three different network scenarios, named \textit{Dumbbell} (D), \textit{Triangle} (T), and \textit{Pyramid Network} (P), whose topologies are reported in Fig.~\ref{fig:networks}.
In all the cases, the number of information flows in the network is $N_{\Phi} \in \{ 2, ..., 6 \}$.}
The capacities of each network element are fixed; specifically, we set $B_l = 50$~Gbps, $C_n^c = 60$~Gbps and $C_n^m = 60$~Gb for the core nodes, $C_n^c = 20$~Gbps and $C_n^m = 20$~Gb for the access nodes.
Hence, we consider that most computing and storage resources are concentrated in the core network.
Concerning the delay, we assume that $\tau_l^p=0.1$~ms, $\forall$ $l \in \mathcal{L}$, and $\tau_n=0.001$~ms, $\forall$ $n \in \mathcal{N}$.
\edit{Although arbitrary, these values are well-aligned with the features of modern network elements.}

To model the information flow requirements, we consider a \gls{mm}~\cite{geyer1992practical} $M_\sigma$ with transition probability matrix $\mathbf{P}_\sigma$ for each slice $\sigma \in \Sigma$.
The model include $N_{\rm state}=10$ states and is designed in such a way that transitions can only occur between adjacent states; mathematically, $\mathbf{P}_{i,j} = 0$ $\forall$ $(i,j): |i-j| > 1$. 
Each state of $M_\sigma$ represents a combination of resources requirements, i.e., a different realization of the vector $\mathbf{r}$. 
Hence, each information flow $\phi \in \Phi_\sigma$ is associated to an independent copy of $M_\sigma$ that changes state at each timeslot $t$, thus altering the vector $\mathbf{r}_\phi$. 
The minimum and maximum value of the resource requirements are summarized in Tab.~\ref{tab:slice_param}.

\begin{table}[h!]
	\centering
	\footnotesize
	\caption{Traffic flow requirements.}
	\label{tab:slice_param}\vspace{-0.2cm}
	\begin{tabular}[c]{@{}c|cccc@{}}
		\toprule
		Service class & Parameter & Range of values & Unit & Sources \\
		\midrule
		& $\eta$ & $0.30 \div 42.5$ & Gbps &~\cite{cominardi2018understanding, 3gpp.23.501}  \\
		\gls{embb}& $c$ & $50 \div 100$ & Gbps &~\cite{sattar2019, chiha2020network} \\
		& $m$ & $50 \div 100$ & Gb &~\cite{sattar2019, chiha2020network}\\
		& $\delta$ & $20$ & ms &~\cite{voigtlander20175g,sachs2018adaptive, jiang2018low, 3gpp.23.203}  \\
		\midrule
		& $\eta$ & $2.08 \div 10$ & Gbps &~\cite{cominardi2018understanding, 3gpp.23.501}  \\
		\gls{urllc}& $c$ & $50-100$ & Gbps &~\cite{sattar2019, chiha2020network} \\
		& $m$ &  $50 \div 100$ & Gb &~\cite{sattar2019, chiha2020network} \\
		& $\delta$ & $1$ & ms &~\cite{voigtlander20175g,sachs2018adaptive, jiang2018low, 3gpp.23.203}\\
		\bottomrule
	\end{tabular}	
\end{table}

To be noted that, in our simulations, the number of flows and their requirements change randomly, so that the aggregate resource requests can exceed the capacity of the network.
In such conditions, some flows will unavoidably experience very low performance; hence, the allocation strategy should decide which flow to penalize, in order to maximize the overall utility.
\edit{Therefore, our system can be exploited to handle critical scenarios where there is a lack of network resources, or to estimate the reliability of a specific set of network slices.}
In the future, we will investigate call-admission control strategies to avoid system overloading. 

\edit{
\begin{table}[h!]
	\centering
	\caption{Agent architectures.}
	\label{tab:learn_param}
	\begin{tabular}{lcccccc}
		\toprule
		& \multicolumn{2}{c}{$\Gamma_\sigma^b$} & \multicolumn{2}{c}{$\Gamma_\sigma^c$} & \multicolumn{2}{c}{$\Gamma_\sigma^m$} \\
		\cmidrule(r){2-3}\cmidrule(l){4-5}\cmidrule(l){6-7}
		Parameter & {Actor} & {Critic}  & {Actor} & {Critic} & {Actor} & {Critic}\\
		\midrule
		Input size & 11 & 11 & 7 & 7 & 7 & 7 \\
		Activation & {ReLU} & {ReLU} & {ReLU} & {ReLU} & {ReLU} & {ReLU}\\
		Hidden size & 12 & 12 & 8 & 8 & 8 & 8\\
		Activation & {ReLU} & {ReLU} & {ReLU} & {ReLU} & {ReLU} & {ReLU}\\
		Hidden size & 6 & 6 & 4 & 4 & 4 & 4\\
		Activation & {Linear} & {Linear} & {Linear} & {Linear} & {Linear} & {Linear}\\
		Output size & 1 & 1 & 1 & 1 & 1 & 1\\
		\bottomrule
	\end{tabular}
\end{table}
}

To train the learning agents, we generate $N_{\rm train}=5 \cdot 10^4$ independent episodes using the same network topology. 
Each episode lasts $N_{\rm slot}=50$ timeslots of $T=0.1$ seconds.
At the beginning of each episode, a random number of information flows is generated; then, each flow $\phi$ is associated with a static route interconnecting its endpoints. 
\edit{Hence, at the end of the episode, the A2C algorithm is used to train the learning agents $\Gamma_\sigma^b$, $\Gamma_\sigma^c$, $\Gamma_\sigma^m$, $\forall$ $\sigma \in \Sigma$.
}
We exploit the \gls{adam} algorithm to optimize the \gls{nn} weights, considering $\zeta_a=10^{-5}$ and $\zeta_c=10^{-5}$ as learning rates of the actor and the critic, respectively. 
We summarize the main settings of the learning architectures in Tab.~\ref{tab:learn_param}, while the other simulation parameters are given in Tab.~\ref{tab:setting}

\edit{
To be noted that the A2C algorithm updates the policies applied by the different agents only at the end of a training episode (i.e., after a predetermined number of time slots).
In the period between two subsequent episodes, the different network elements collect local observations in a central database, which is then used to carry out the training phase.
Hence, the new versions of the learning agents is shared within the network, and the resource allocation strategy is updated consequently. 
In a practical scenario, the described process can be performed offline and, therefore, does not have any delay requirements.
Besides, the exchange of a control traffic among network elements is required by the \gls{sdn} and \gls{nfv} paradigms, which are becoming more and more widespread in modern networks.
Therefore, the communication within the learning units should not represent a limiting factor for the practical implementation of our approach.}

\begin{table*}[t!]
	\centering
	\footnotesize
	\caption{Simulation settings.}
	\label{tab:setting}
	\begin{tabular}{@{}ccc|ccc@{}}
		\toprule
		Parameter & Value & Description & Parameter & Value & Description \\ \midrule
		$B_l$ & $50$ Gbps & Link rate capacity & $C_n^c$ & $\{10, 20, 30, 60\}$ Gbps & Node computational capacity \\
		$C_n^m$ & $\{10, 20, 30, 60\}$ Gb & Node memory capacity & $N_{\rm state}$ & 10 & State number \\
		$\tau_n$ & $0.001$ ms & Node routing delay & $\tau_l^p$ & $0.1$ ms & Link propagation delay \\
		$N_{\rm train}$ & $\{3, 5\} \cdot 10^4$ & Training episodes & $N_{\rm test}$ & $500$ & Testing episodes  \\
		$N_{\rm transfer}$ & $2 \cdot 10^4$ & Transfer learning episodes & $N_{\Phi}$ & $\{2,...,6\}$ & Number of flows \\		
		$\zeta_a$ & $10^{-5}$ & Actor learning rate &  $\zeta_c$ & $10^{-5}$ & Critic learning rate    \\
		$\lambda$ & $0.9$ & Discount factor &  $\kappa$ & $10^{-4}$ & Entropy weight \\
		$T$ & $0.1$ s & Timeslot duration &  $N_{\rm slot}$ & $50$ & Timeslots per episode \\
		$\{\alpha_{\eta}, \alpha_{c}, \alpha_{m}, \alpha_{\delta} \}$ & $\{0.25, 0.25, 0.25, 0.25\}$ & \gls{embb} performance weights & $\{\gamma_0, \gamma_1\}$ & $\{0.1, 1\}$ & Reward weights \\
		\midrule
	\end{tabular}
\end{table*}

\subsection{Results}\label{sec:result}

We consider two versions of our learning system, one trained in the Dumbbell Network (DRL-D) and the other in the Triangle Network (DRL-T).

\begin{figure}[t!]
	\centering
	\begin{subfigure}{0.9\linewidth}
		\centering
		\includegraphics[width=\linewidth]{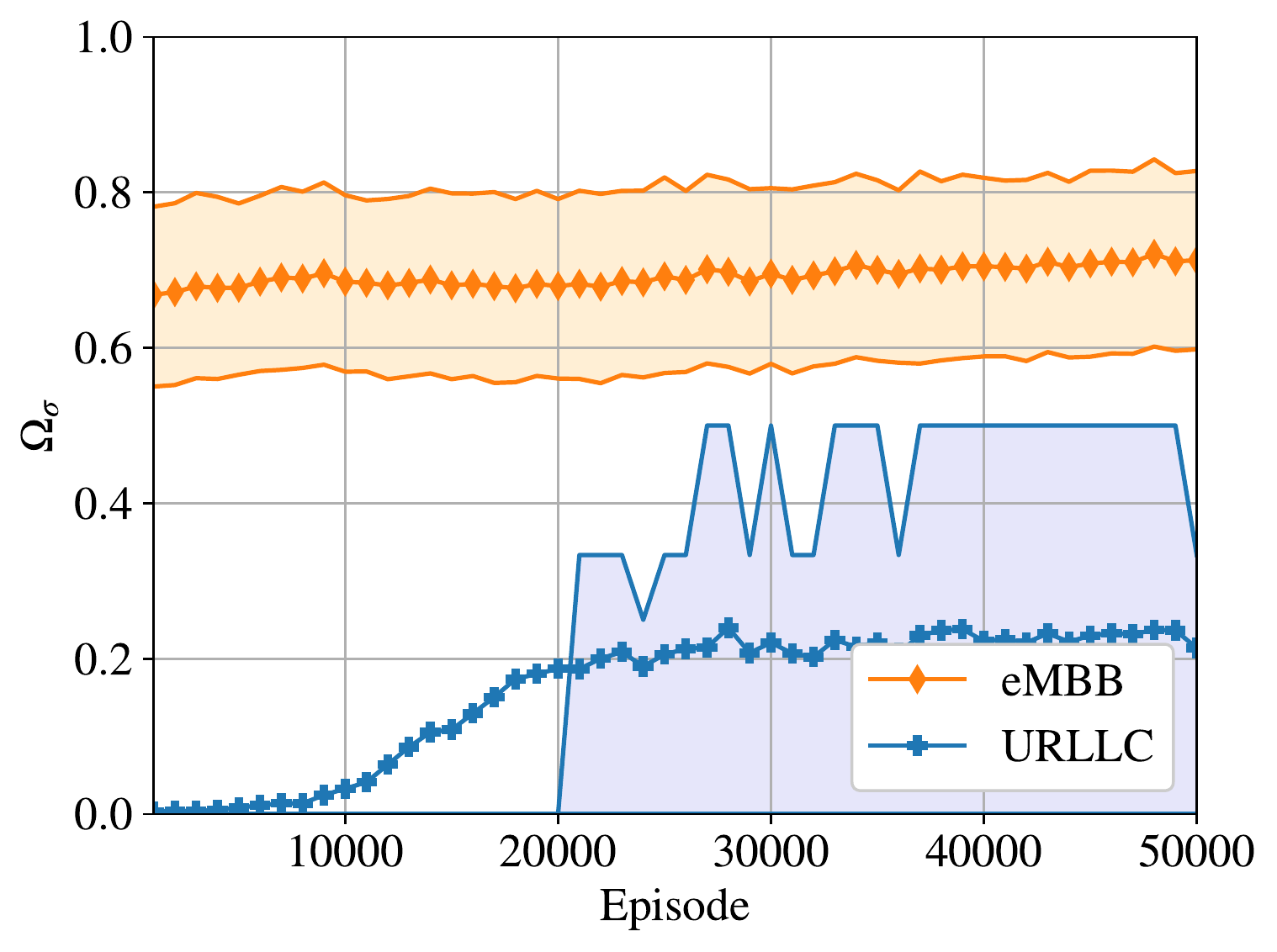}
		\caption{Dumbbell Network.}
		\label{fig:embb_train}
	\end{subfigure}	
	\begin{subfigure}{0.9\linewidth}
		\centering
		\includegraphics[width=\linewidth]{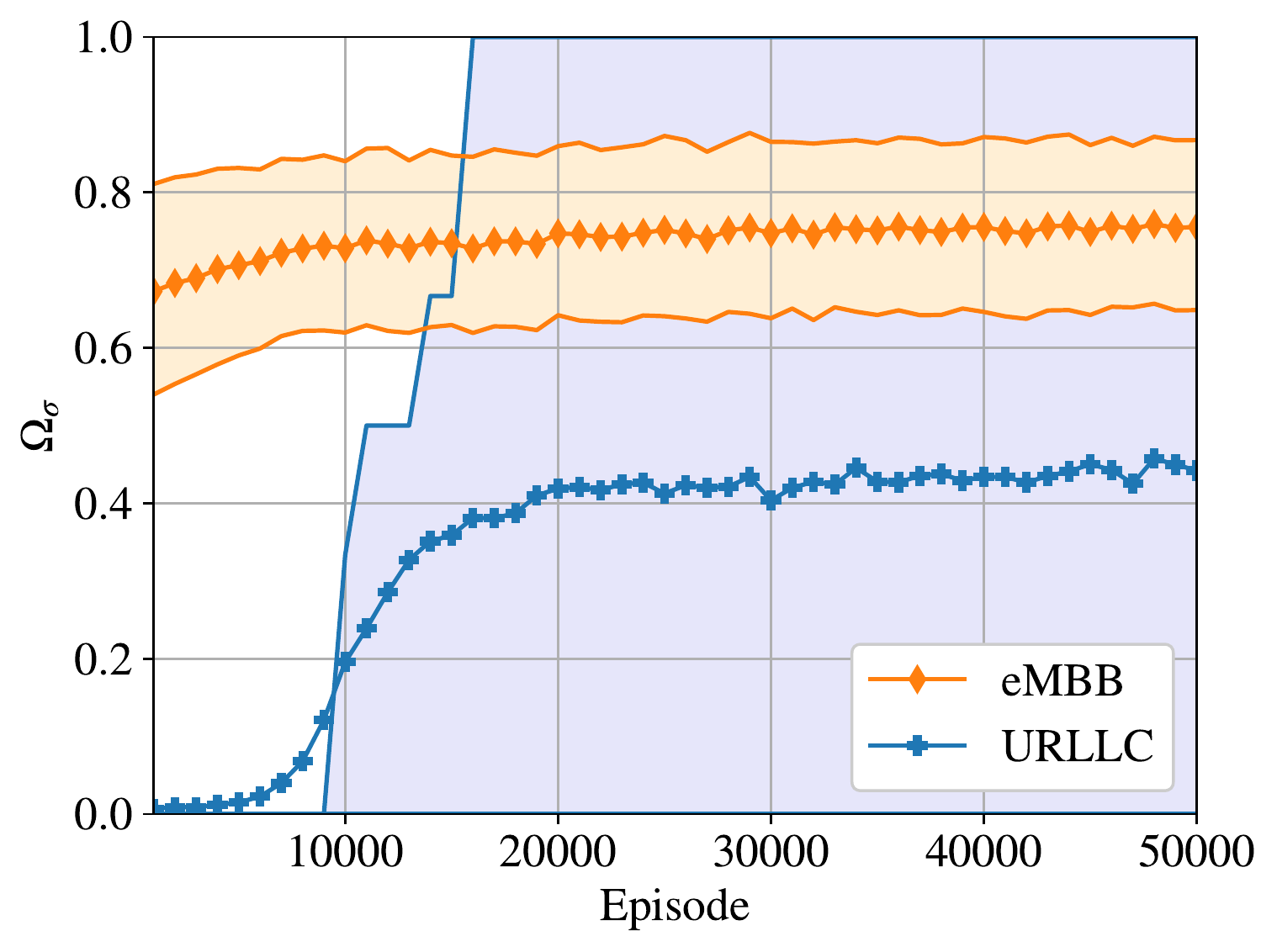}
		\caption{Triangle Network.}
		\label{fig:urllc_train}
	\end{subfigure}
	\caption{Training phase.}
	\label{fig:train}
\end{figure}

\edit{In Fig.~\ref{fig:train} we plot the utility (for the \gls{embb} and \gls{urllc} slices) obtained during the training phase in the two scenarios.
that the performance of \gls{embb} slices ($\Omega_e$) increases slowly, but smoothly during the training phase, ranging from 0.6 to 0.8 in 50\% of the cases.
Conversely, the performance of \gls{urllc} slices ($\Omega_u$) remains very low at the beginning of the training phase and suddenly increases after a certain number of episodes. 
This means that the agents need more time to learn how to address the \gls{urllc} requirements.}
In particular, at the end of the training phase in the Triangle Network, $\Omega_u$ spans the full range of possible values, which indicates that the \gls{urllc} flows either get maximum or zero reward in accordance the step-like shape of their performance function.
\edit{A similar phenomenon occurs in the Dumbbell Network where, however, the \gls{urllc} utility is lower due to the fewer available resources.}

\begin{figure}[t!]
	\centering
	\begin{subfigure}{0.9\linewidth}
		\centering
		\includegraphics[width=\linewidth]{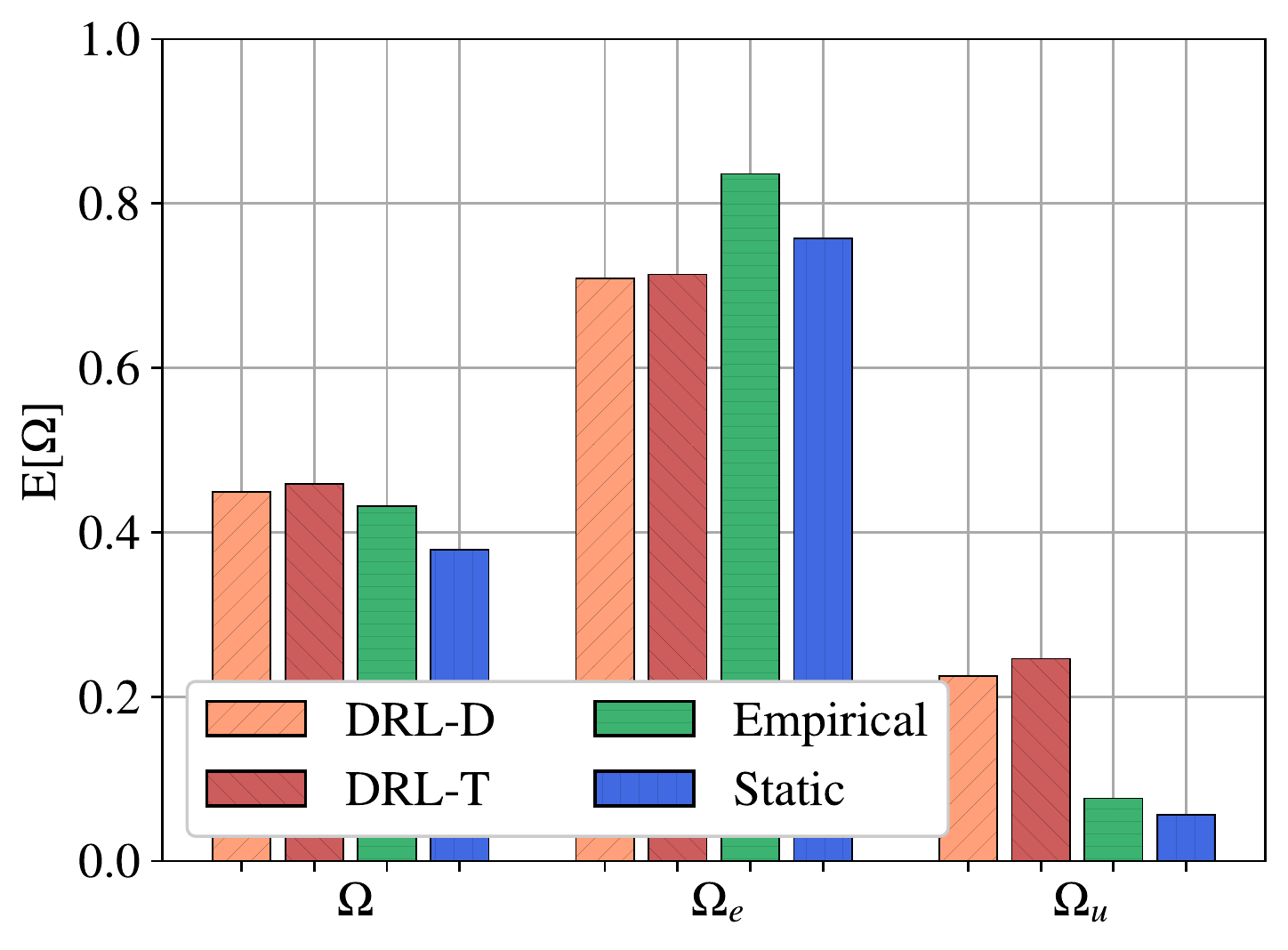}
		\caption{Dumbbell Network.}
		\label{fig:dumbell_expected}
	\end{subfigure}	
	\begin{subfigure}{0.9\linewidth}
		\centering
		\includegraphics[width=\linewidth]{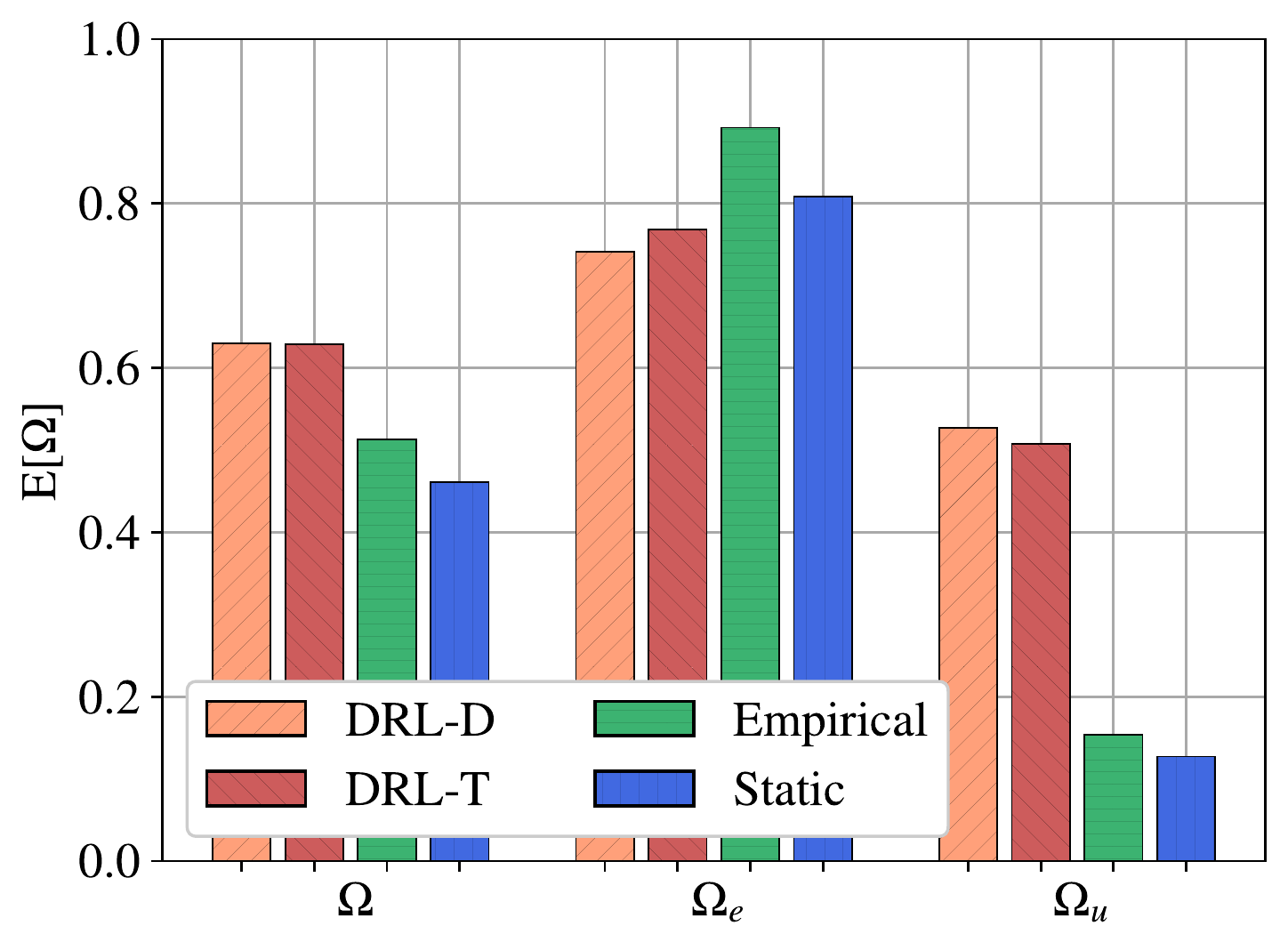}
		\caption{Triangle Network.}
		\label{fig:triangle_expected}
	\end{subfigure}
	\caption{Expected utility.}
	\label{fig:expected_utility}
\end{figure}

\edit{To test the performance of our strategies, we carry out additional $N_{\rm test}=300$ episodes.}
In Fig.~\ref{fig:expected_utility}, we report the expected utility achieved by each slice, and by the whole system (see \eqref{eq:slice_perf} and \eqref{eq:system_perf}), in the Dumbbell and Triangle Network for all the considered strategies.
We can notice that, in both the scenarios, the empirical algorithm tends to favor \gls{embb} slices at the expenses of \gls{urllc} flows, whose expected performance is always lower than $0.2$.
The static strategy behaves similarly but provides lower $\Omega_e$ than the empirical algorithm.
In contrast, DRL-D and DRL-T \edit{slightly reduce the performance of the \gls{embb} services in order to double the fraction of satisfied \gls{urllc} flows, increasing the total utility.}

\begin{figure}[t!]
	\centering
	\includegraphics[width=0.9\linewidth]{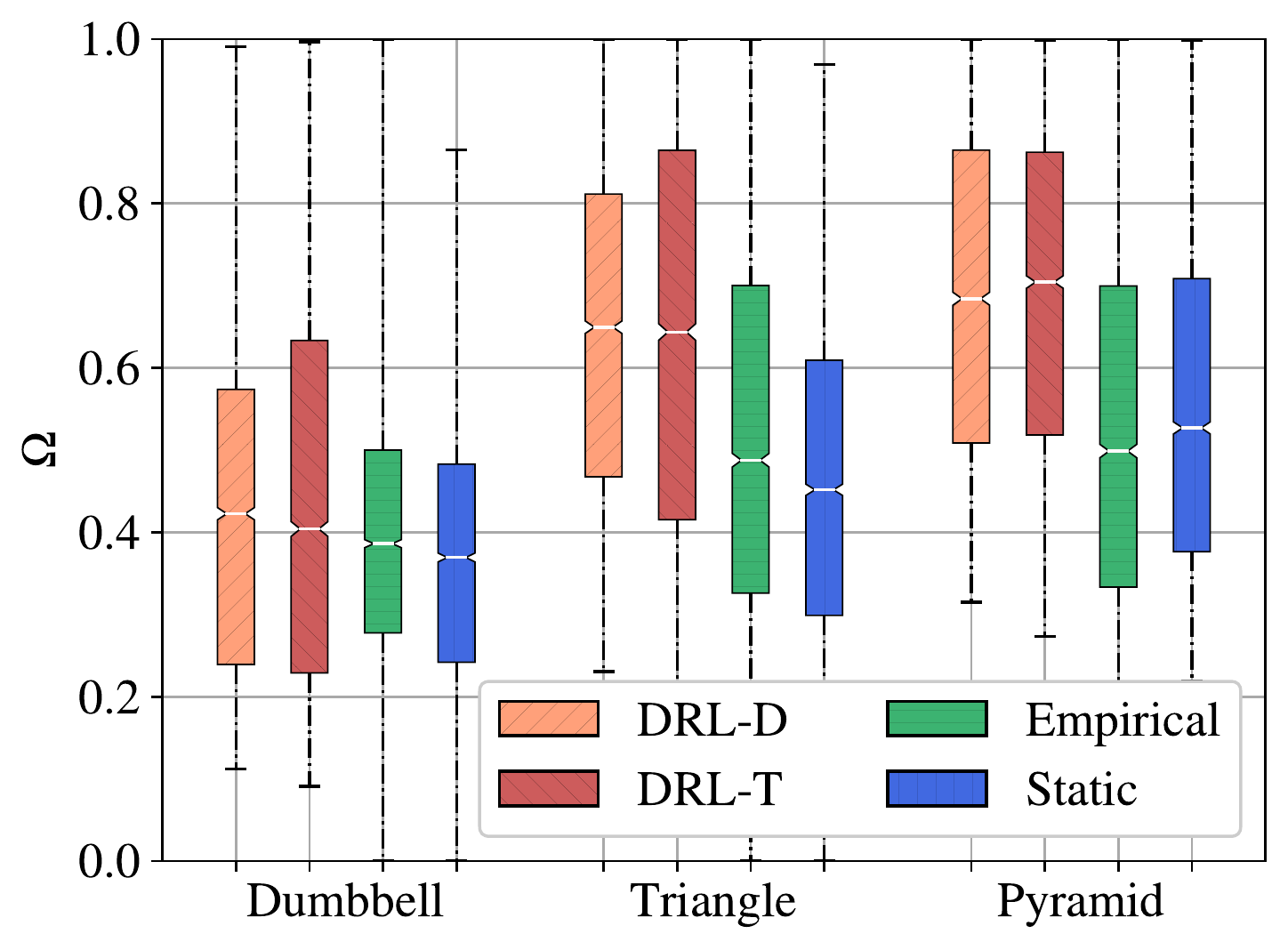}
	\caption{Utility distribution.}
	\label{fig:icc_box}
\end{figure}

In Fig.~\ref{fig:icc_box} we analyze the distribution of the system utility for all the different strategies and network scenarios.
We adopt the boxplot representation, where the white line at the center of the box is the median of the distribution, while the box edges represent the $25$th and the $75$th percentile, respectively. 

It is clear that the static strategy always yield to the worst performance: this is because it does not handle the variability of service demands, which is very critical for the \gls{urllc} traffic flows, whose performance goes to zero as soon as one of the requirements is not satisfied. 
The empirical algorithm works better but it is still outperformed by the \gls{drl} strategies, both when considering the median and the $25$th and $75$th percentiles of $\Omega$. 
In the Dumbbell Network, DRL-D and DRL-T ensure that $\Omega > 0.45$ in almost $50\%$ of the test episodes, with a $10\%$ gain over the empirical algorithm.
In the Triangle and Pyramid scenario, the lack of network resources is less striking and, consequently, the performance of all the strategies increases.
Using the empirical algorithm, $50\%$ of the test episodes experience $\Omega > 0.5$ while, with the \gls{drl} strategies, this threshold is raised to $0.65$ and $0.7$, respectively. 

We observe that DRL-D and DRL-T achieve similar results in all the testing scenarios, \edit{including the Pyramid Network, which is a different environment from those seen during the training.}
This means that our architecture can suit multiple network topologies without the need of an additional learning phase.
At the same time, we expect that a more specific training can further improve the behavior of the local controllers; in the rest of the section, we will show how to leverage the transfer learning paradigm to this purpose. 

\subsection{Transfer Learning}

\begin{figure}[t!]
	\centering
	\begin{subfigure}{0.9\linewidth}
		\centering
		\includegraphics[width=\linewidth]{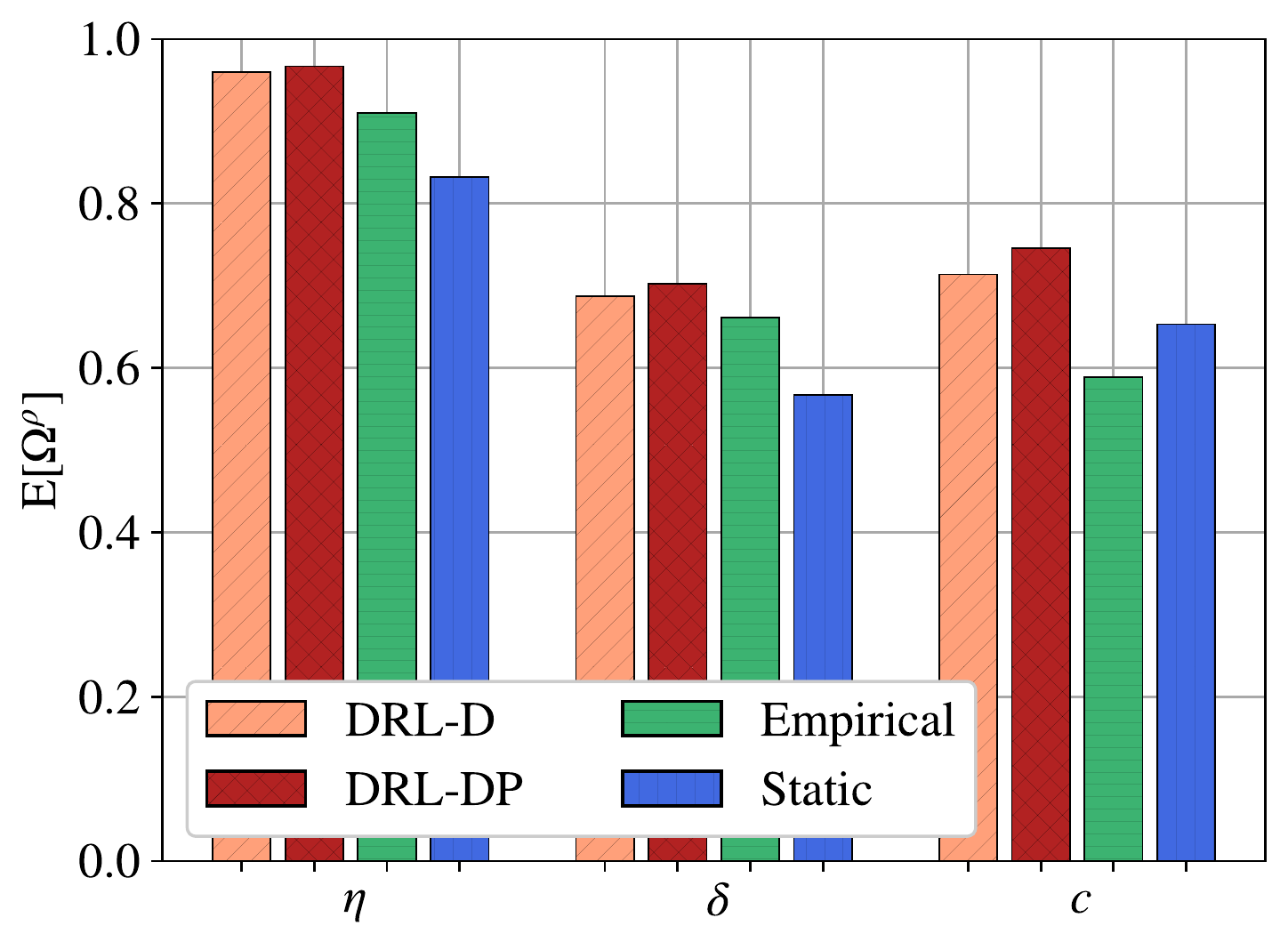}
		\caption{Pyramid Network.}
		\label{fig:big_bar_transfer}
	\end{subfigure}	
	\begin{subfigure}{0.9\linewidth}
		\centering
		\includegraphics[width=\linewidth]{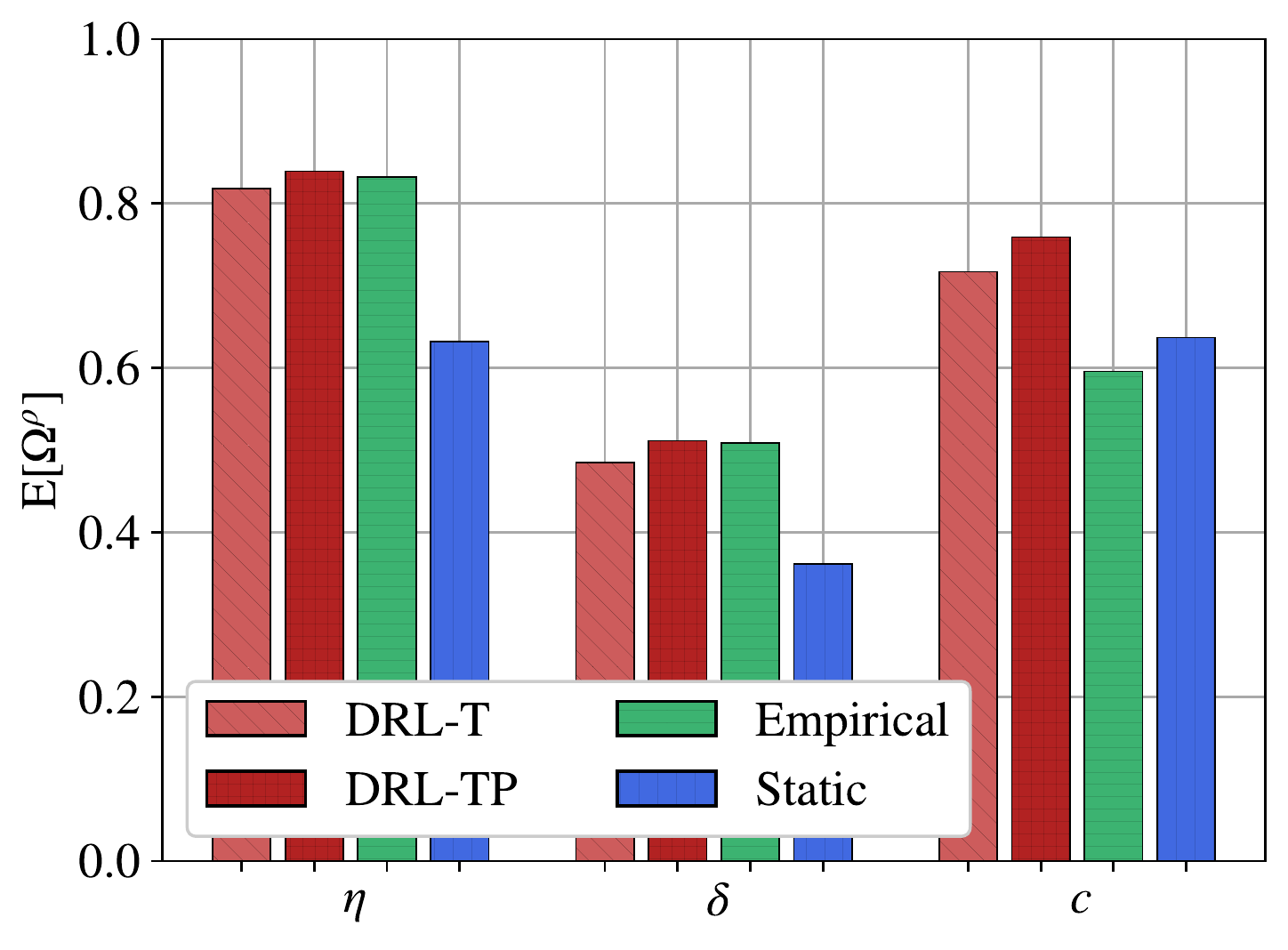}
		\caption{Pyramid+ Network.}
		\label{fig:bbig_bar_transfer}
	\end{subfigure}
	\caption{Expected resource utility.}
	\label{fig:bar_transfer}
\end{figure}

\edit{Transfer learning aims at speeding up the training of a \gls{ml} algorithm in a certain scenario by exploiting the structure learned by other \gls{ml} algorithms trained in similar scenarios. 
In what follows, we exploit this technique to adapt our \gls{drl} strategy to different network topologies and traffic loads.}
Hence, we consider two additional learning architectures that we named DRL-DP and DRL-TP. 
These systems are first trained for $N_{\rm train} = 3 \times 10^4$ episodes in the Dumbbell and Triangle Network, respectively, according to the framework described in Sec.~\ref{sec:drl}. 
Then, we perform an additional training of $N_{\rm transfer} = 2 \times 10^4$ episodes in the Pyramid Network.

During the additional training phase, each controller is updated with experience related to the network element it is associated to.
For instance, a controller $\Gamma$ designed to manage slice $\sigma$ in a link $l$ is trained using only the state-action pairs for link $l$ and information flows $\phi \in \Phi_\sigma$. 
\edit{Therefore, each controller has to deal with a new scenario with different characteristics than the original one.
Particularly, the training operations can be performed online in each network element, without involving the central manager shown in Fig.~\ref{fig:architecture}.
Hence, this stage does not require any communication within the network and can be executed after the learning architecture have been deployed in a real scenario.}
From a practical perspective, the described framework makes local controllers learn how to carry out more precise actions, thus increasing the overall utility.
\edit{The drawback is that each controller will be able to operate only in a specific location and, therefore, the learned strategy cannot be implemented in different network topologies.}

\edit{The transfer learning stage is repeated two times, considering different configurations for the \gls{urllc} services. 
First, we implement the same statistics presented in Tab.~\ref{tab:slice_param}: in this case, the throughout required by each \gls{urllc} flow is in $[2.08, 10]$~Gbps. 
Then, we double the rate requirements of the \gls{urllc} slice (whose range of values becomes $[4.16, 20]$~Gbps) to asses the ability of our strategy to adapt to new service specifications.
In particular, we denote by Pyramid+ the scenario in which the required throughput of the \gls{urllc} flows is increased.}

\begin{figure}[t!]
	\centering
	\includegraphics[width=0.9\linewidth]{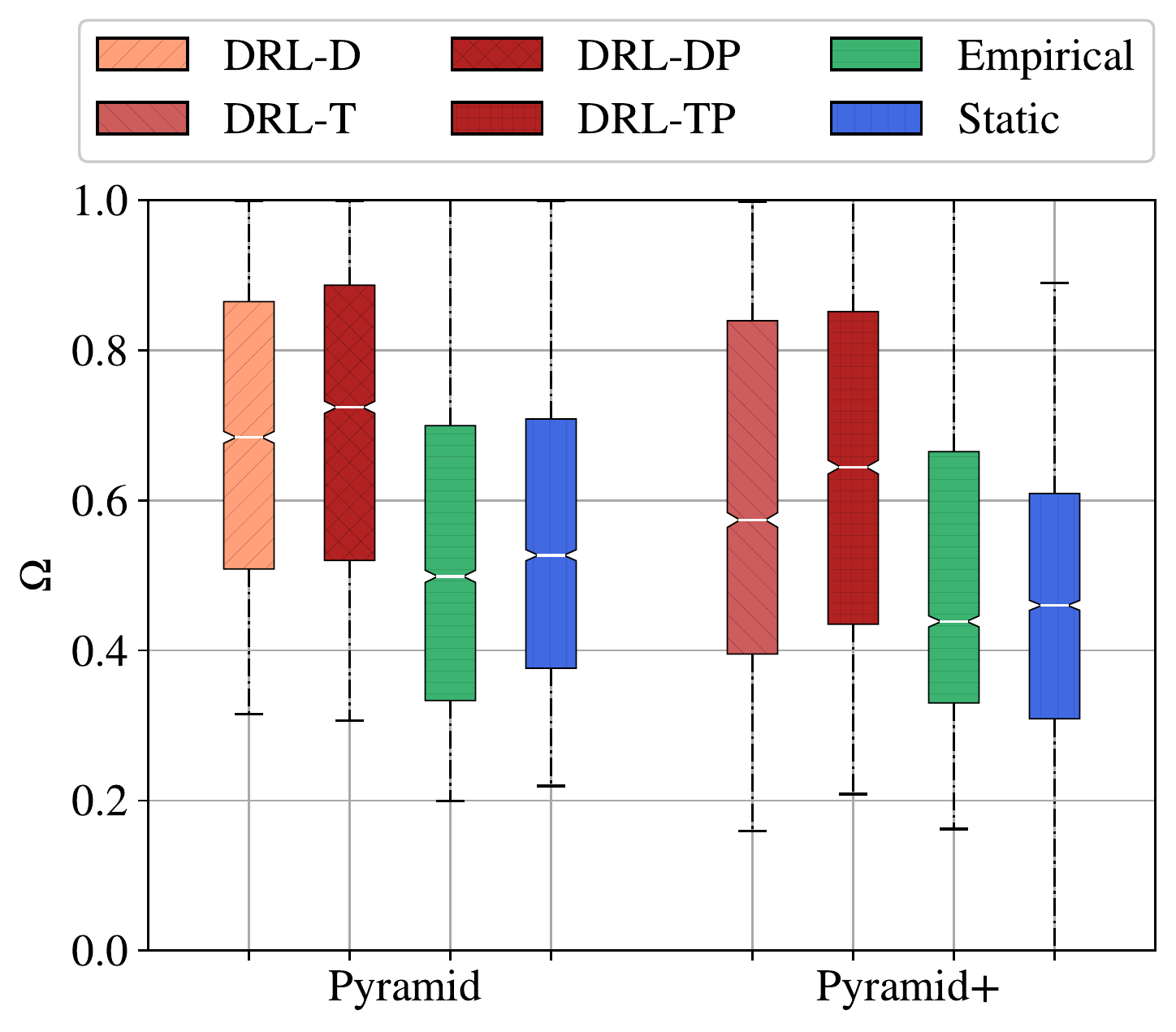}
	\caption{Utility distribution.}
	\label{fig:boxplot_transfer}
\end{figure}

\edit{In Fig.~\ref{fig:big_bar_transfer}, we represent the expected performance obtained in the Pyramid Network by DRL-D and DRL-DP, and the benchmark strategies, while considering specific network resources.
Hence, instead of using $\Omega$ to asses the system utility, we consider $\Omega^\rho$, defined in \eqref{eq:slice_res_perf}. 
As expected, the algorithms obtained by exploiting the transfer learning approach perform better than those trained on the other networks.
For instance, using the DRL-DP strategy, the expectation of $\Omega^c$ increases by more than $5 \%$.}

\edit{In Fig.~\ref{fig:bbig_bar_transfer}, we show the outcomes obtained in the Pyramid+ network scenario.
We first observe that all the strategies yield to a lower performance since the new \gls{urllc} requirements are more difficult to fulfill. 
The empirical strategy slightly outperforms DRL-T for what concerns the throughput and delay requirements.
Nevertheless, DRL-T still provides good results, which means that the \gls{drl} approach is resilient to different traffic loads. 
Also in this case, transfer learning proves to be a worthwhile approach since DRL-TP outperforms the other algorithms in all three key performance indicators.
}

\edit{In Fig.~\ref{fig:boxplot_transfer}, we report the distribution of the system utility in the Pyramid and Pyramid+ scenarios.
We can appreciate how the the algorithms given by transfer learning always ensure the best utility, both considering the median and the percentiles of $\Omega$.
In particular, DRL-DP and DRL-TP outperform DRL-D and DRL-T, respectively, despite the total number of training episodes is the same for all the considered strategies.
}

\begin{figure}[t!]
	\centering
	\includegraphics[width=0.9\linewidth]{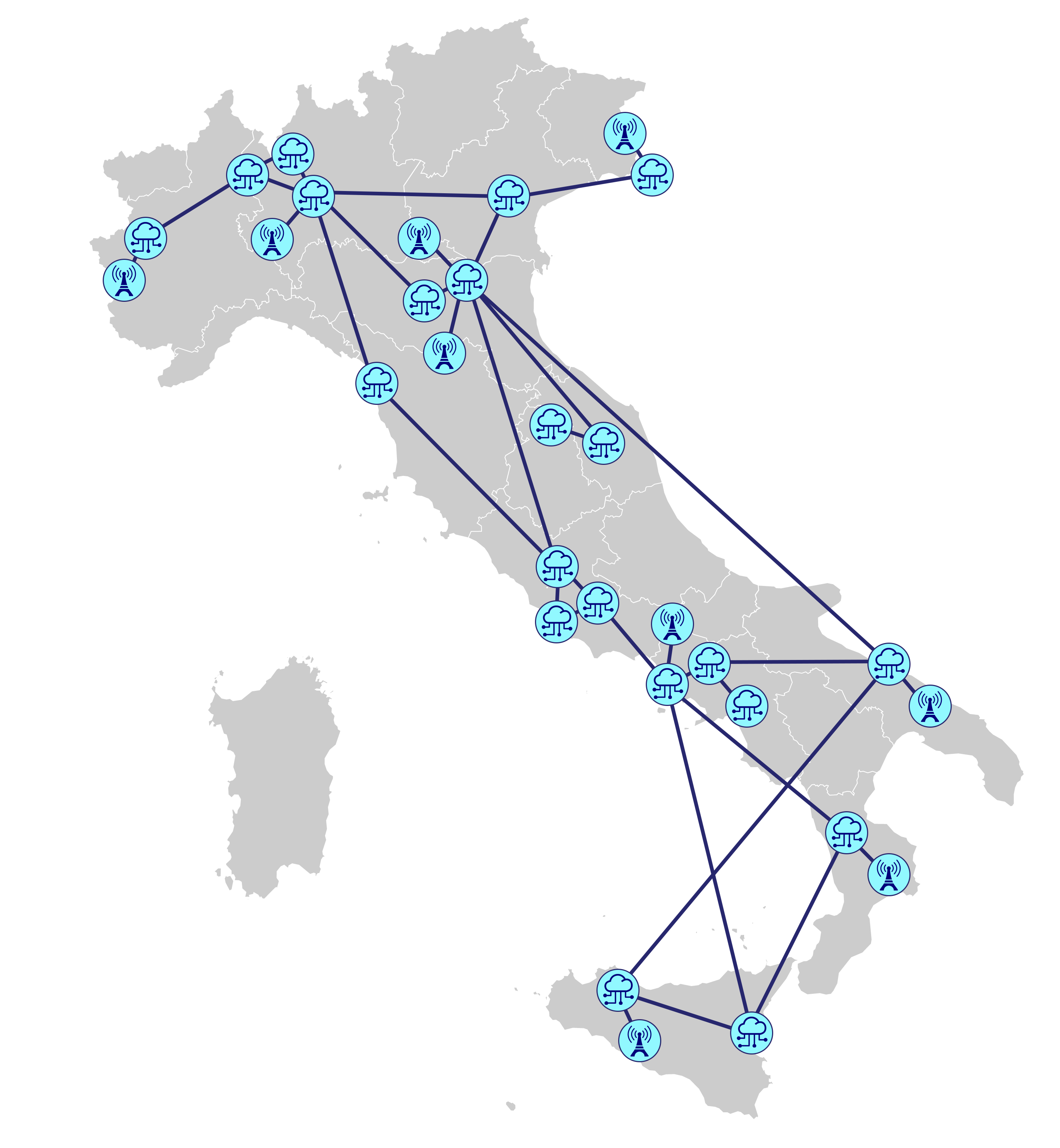}
	\caption{GARR Network (Italy).}
	\label{fig:garr_topology}
\end{figure}

\begin{figure}[t!]
	\centering
	\includegraphics[width=0.9\linewidth]{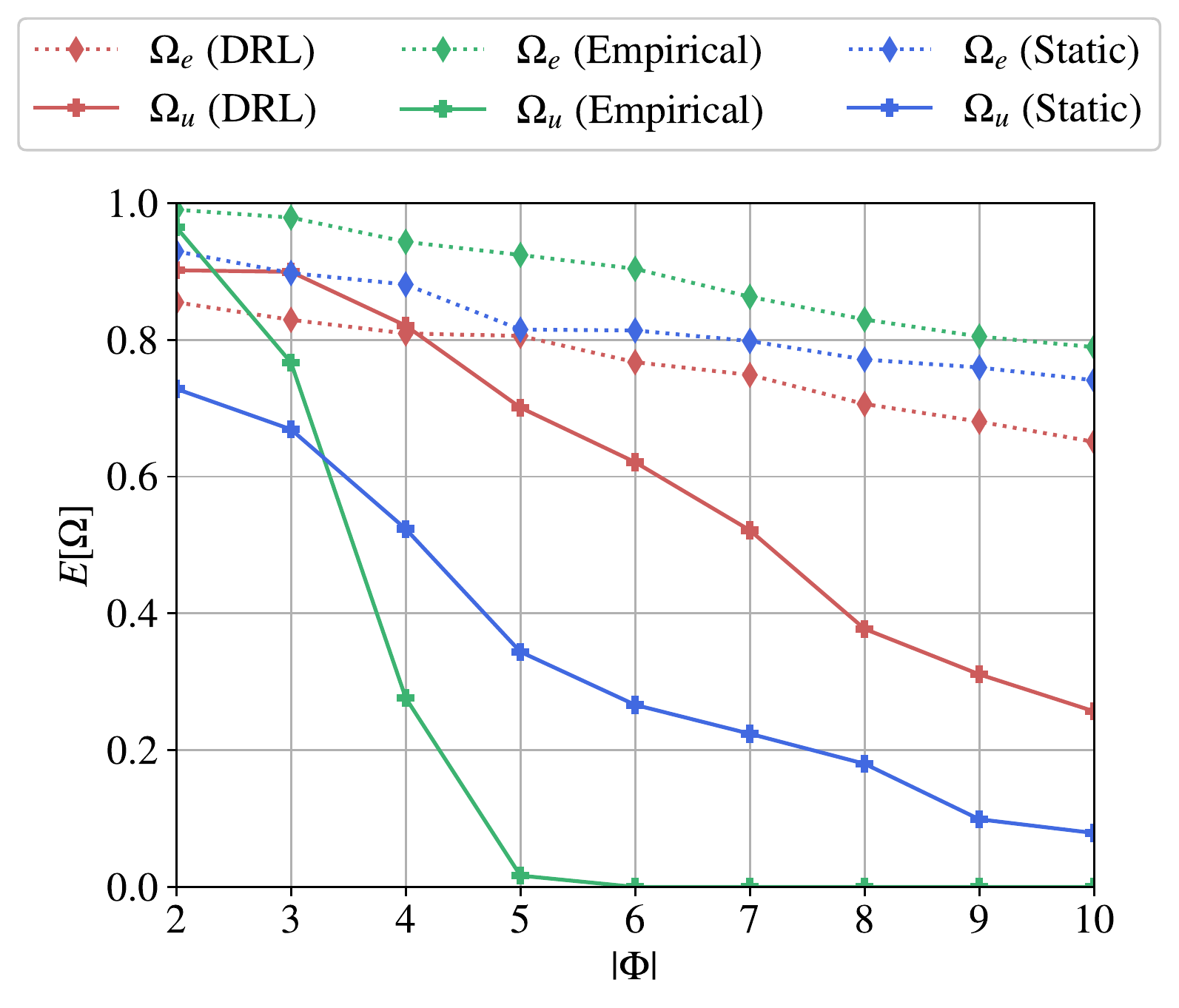}
	\caption{Expected utility vs flows' number.}
	\label{fig:omega_flow}
\end{figure}

\subsection{Practical Implementation}

\edit{
To provide an overview on how implement our system in a realistic scenario, we consider a section of the GARR high capacity network~\cite{garr}, which is the infrastructure interconnecting the main Italian universities and research centers. 
As depicted in Fig.~\ref{fig:garr_topology}, the considered network topology includes $19$ core nodes, $10$ edge nodes and $40$ links.
We assume that each link is provided with $B_l = 50$~Gbps of bandwidth capacity; besides, we set the computational and memory capacities of the core nodes to $C_n^c = 30$~Gbps and $C_n^m = 30$~Gb, respectively, and those of the edge nodes to $C_n^c = 10$~Gbps and $C_n^m = 10$~Gb, respectively.
Finally, we relax the delay requirements of the \gls{embb} and \gls{urllc} services, which are set to $100$ ms and $5$ ms, respectively.
We make this choice to deal with the increased length of the routing paths, which results in a higher propagation delay for the flows.}

\edit{To implement our DRL strategy in the new scenario, we carry out two consecutive learning stages: first, we train our system for $2\times10^4$ episodes using a centralized learning approach (as done for the DRL-D and DRL-T strategies).
We recall that this phase can be achieved offline, i.e., without time constraints, using all the experience gathered all the network locations. 
Then, we carry out an additional training of $10^4$ episodes, defining a specific tuple of learning agents for each network element (as done for the DRL-DP and DRL-TP strategies).
As already stated, this stage is performed online and does not require any communication within the network. 
}

In Fig.~\ref{fig:omega_flow}, we plot the expected performance of the \gls{embb} and  \gls{urllc} slices (i.e., E[$\Omega_e$] and E[$\Omega_u$]) as a function of the number of information flows in the GARR scenario. 
When considering the \gls{embb} flows, all the strategies have a similar behavior since $E[\Omega_e]$ declines gradually as the cardinality of $\Phi$ increases. 
Instead, $E[\Omega_u]$ deceases very quickly: it is maximized when $|\Phi| = 2$ and approaches $0$ as more flows are initialized over the network. 

In particular, the empirical algorithm outperforms the other strategies when the number of information flows is limited ($|\Phi| \leq 3$). 
Beyond this point, it becomes more convenient to maintain a static allocation of the system resources to prevent the degradation of $E[\Omega_u]$.
However, both the benchmarks are outperformed by the learning-based approach, \edit{which has learned to give priority to the \gls{urllc} services, ensuring a better total utility}. 
In particular, using the empirical and the static strategies, $E[\Omega_u]$ falls below $0.3$ for $|\Phi|=6$, while DRL ensures $E[\Omega_u] > 0.6$ in the same conditions.

\section{Conclusion}\label{sec:conclusion}

In this work, we investigated the potentials of \gls{drl} to orchestrate network resources in a \gls{ns} scenario. 
Specifically, we developed a distributed \gls{drl} system where different units interact to meet the resource demands of multiple information flows. 
The training of such an architecture was undertaken following an \gls{a2c} approach, \edit{which enables an online training phase and allows our system to dynamically adapt to different working conditions}.

By means of simulations, we showed that the designed strategy can consistently improve the management of network resources, \edit{especially when the system complexity, both in terms of network topology and service heterogeneity, increases.}
In particular, our approach makes it possible to double the number of \gls{urllc} flows supported by the network, without significantly degrading the performance of the \gls{embb} flows, and can suit different network topologies without the need for additional training.
Besides, transfer learning can be used to further improve the behavior of the learning agents, thus increasing the overall utility at the cost of a reduced adaptability of the agents to different network topologies. 

As part of future work, we are interested in extending our \gls{ns} model by considering more slice classes with different specifications. 
Particularly, we want to test our framework with real communication traces, with the aim of identifying potential limits.
Finally, we will investigate the possibility of introducing additional learning units, which are trained to coordinate the local controllers of our architecture\edit{, e.g., by varying the routing paths of the traffic flows.}

\appendix

In what follows, we derive equation \eqref{eq:const_15}, which determines the average queuing time $\tau_{l,\phi}^q(t)$ of  $\phi$ in link $l$ during the time slot $t$.
Let $T^*$ be the minimum between the slot boundary $T$ and the time at which the queue empties, i.e.,

\begin{equation}
	T^*=\min \left(T, \frac{D_{l,\phi}(t-1)}{b_{l,\phi}(t)-b^i_{l,\phi}(t)}\right).
\end{equation}

Furthermore, let $\mathcal{D}_{l,\phi}(t,u)$ denote the residual backlog of flow $\phi$ in link $l$, $u$ seconds after the beginning of timeslot $t$.
Therefore, for any $u\in [0,T^*]$ we have $\mathcal{D}_{l,\phi}(t,u)=D_{l,\phi}(t) -u (b_{l,\phi}(t)-b^i_{l,\phi}(t)) $, while $\mathcal{D}_{l,\phi}(t,u)=0$ for $u\in (T^*, T]$.
Now, the queuing delay experienced by the incoming flow at time $u \in [0,T]$ is zero if the queue is empty, and otherwise equal to $
\delta(u,t) = \frac{\mathcal{D}_{l,\phi}(u,t)}{b_{l,\phi}(t)}  
$.  
The average delay over the timeslot is hence

\begin{align}
\tau_{l,\phi}^q(t) =& \frac{1}{T}\int_{0}^T \delta(u,t) \mathrm{d}u =\frac{1}{T}\int_{0}^{T^*} \frac{\mathcal{D}_{l,\phi}(u,t)}{b_{l,\phi}(t)} \mathrm{d}u   \\
=&\frac{2T^*D_{l,\phi}(t) - {T^*}^2( b_{l,\phi}(t)- b^i_{l,\phi}(t))}{2T b_{l,\phi}(t)} .
\label{xxx}
\end{align}
For $T^*=T$, we obtain
\begin{equation}
	\tau_{l,\phi}^q(t) =
	\frac{2 D_{l,\phi}(t) - T( b_{l,\phi}(t)- b^i_{l,\phi}(t))}{2 b_{l,\phi}(t)} .
\end{equation}
For $T^*<T$, instead, we have ${T^*}( b_{l,\phi}(t)- b^i_{l,\phi}(t)) = D_{l,\phi}(t)$, so that \eqref{xxx} yields 
\begin{equation}
	\tau_{l,\phi}^q(t) = \frac{T^* D_{l,\phi}(t) }{2T b_{l,\phi}(t)} < \frac{D_{l,\phi}(t) }{2b_{l,\phi}(t)} .
\end{equation}

Recalling that $b_{l,\phi}^o (t)$ used in \eqref{eq:const_15} is defined as the minimum between $b_{l,\phi}(t)$, and $\frac{2D_{l,\phi}(t-1)}{T}+b_{l,\phi}^i(t)$, we can see that \eqref{eq:const_15} is indeed a compact expression for $\tau_{l,\phi}^q(t) $, provided that it is approximated by its upper bound when $T^*<T$.

\bibliographystyle{IEEEtran}
\bibliography{bibliography.bib}

\end{document}

%% file: acronyms.tex
\newacronym{sla}{SLA}{Service Level Agreement}
\newacronym{slo}{SLO}{Service Level Objective}
\newacronym{qos}{QoS}{Quality of Service}
\newacronym{qoe}{QoE}{Quality of Experience}
\newacronym{embb}{eMBB}{enhanced Mobile BroadBand}
\newacronym{urllc}{URLLC}{Ultra Reliable Low Latency Communication}
\newacronym{mmtc}{mMTC}{massive Machine Type Communication}
\newacronym{sdn}{SDN}{Software Defined Networking}
\newacronym{nfv}{NFV}{Network Function Virtualization}
\newacronym{ns}{NS}{Network Slicing}
\newacronym{drl}{DRL}{Deep Reinforcement Learning}
\newacronym{iot}{IoT}{Internet of Things}
\newacronym{bbu}{BBU}{BaseBand Unit}
\newacronym{inp}{InP}{Infrastructure Provider}
\newacronym{bs}{BS}{Base Station}
\newacronym{ml}{ML}{Machine Learning}
\newacronym{ac}{AC}{Actor Critic}
\newacronym{a2c}{A2C}{Advantage Actor Critic}
\newacronym{nn}{NN}{Neural Network}
\newacronym{slqp}{SLQP}{Sequential Linear-Quadratic Programming}
\newacronym{ga}{GA}{Genetic Algorithm}
\newacronym{relu}{ReLU}{Rectifier Linear Unit}
\newacronym{rl}{RL}{Reinforcement Learning}
\newacronym{mdp}{MDP}{Markov Decision Process}
\newacronym{mm}{MM}{Markov Model}
\newacronym{adam}{Adam}{Adaptive moment estimation}
\newacronym[longplural={Points of Presence}
]{pop}{PoP}{Points of Presence}
\newacronym{mec}{MEC}{Mobile Edge Computing}
\newacronym{ran}{RAN}{Radio Access Network}
\newacronym{cn}{CN}{Core Network}